\documentclass[12pt,fleqn,twoside]{article}

\usepackage{times}
\usepackage[german,english]{babel}
\usepackage{dbairep}

\foreignreport

\authornames{
Wolfgang Slany
}

\published{Online version of this paper:
http://www.dbai.tuwien.ac.at/ftp/papers/slany/dbai-tr-99-34.ps.gz}

\abstract{
\noindent We consider combinatorial avoidance and achievement games
based on graph Ramsey theory: The players take turns in coloring still
uncolored edges of a graph $G$, each player being assigned a distinct
color, choosing one edge per move. In avoidance games, completing a
monochromatic subgraph isomorphic to another graph $A$ leads to
immediate defeat or is forbidden and the first player that cannot move
loses. In the avoidance$^+$ variants, both players are free to choose
more than one edge per move. In achievement games, the first player
that completes a monochromatic subgraph isomorphic to $A$ wins.
\citet{Erdos73} were the first to identify some tractable subcases of
these games, followed by a large number of further studies. We
complete these investigations by settling the complexity of all
unrestricted cases: We prove that general graph Ramsey avoidance,
avoidance$^+$, and achievement games and several variants thereof are
\pspacec. We ultra-strongly solve some nontrivial instances of graph
Ramsey avoidance games that are based on symmetric binary Ramsey
numbers and provide strong evidence that all other cases based on
symmetric binary Ramsey numbers are effectively intractable.
\vfill
}

\infsysrr{DBAI-TR-99-34}
\date{November 5, 1999}

\newcommand{\boxcaption}[2]{\begin{center}\parbox{.9\columnwidth}{\caption{#1}\label{#2}}\end{center}}

\newcommand{\rem}[1]{}
\newcommand{\hidden}[2]{}

\usepackage{epsfig}
\usepackage[numbers]{natbib}
\usepackage{latexsym}

\newtheorem{theo}{Theorem}
\newtheorem{lem}[theo]{Lemma}
\newtheorem{corol}[theo]{Corollary}
\newtheorem{defn}{Definition}[section]
\newtheorem{claim}{Claim}[section]
\newtheorem{obs}{Observation}[section]
\newtheorem{fact}[obs]{Fact}
\newtheorem{conj}{Conjecture}
\newtheorem{open}{Open Problem}

\newcommand{\sproof}[1]{\subsection{Proof of Theorem~\ref{#1}}}
\newcommand{\scor}[1]{\subsection{Proof of Corollary~\ref{#1}}}
\def\proof{\relax{\noindent\bf Proof.\ }}
\def\halmos{\hspace*{1em}{\rule[-0.5mm]{1.5mm}{3mm}}}
\def\halmoseol{\hspace*{\fill}\halmos}

\newcommand{\And}{\wedge}
\newcommand{\Or}{\vee}
\newcommand{\de}{\stackrel{\mathrm{def}}{=}}
\newcommand{\rde}{\stackrel{\mathrm{redef}}{=}}

\newcommand{\N}{{{\sf l} \kern -.10em {\sf N} }}

\newcommand{\p}{{\bf P}}

\newcommand{\np}{{\bf NP}}

\newcommand{\pspace}{{\bf PSPACE}}
\newcommand{\pspacec}{\pspace{}-complete}

\newcommand{\exptimec}{{\bf EXPTIME}-complete}
\newcommand{\expspace}{{\bf EXPSPACE}}
\newcommand{\logspace}{{\bf LOGSPACE}}

\newcommand{\piptc}{$\mathbf{\Pi}^{\bf P}_2$-complete}
\newcommand{\pipt}{$\mathbf{\Pi}^{\bf P}_2$}

\newcommand{\lcm}{\mathrm{lcm}}
\newcommand{\setcomma}{\,,\,}
\newcommand{\union}{\:\cup\:}

\newcommand{\ramsey}{\mbox{$\mathrm{Ramsey}$}}
\newcommand{\gramsey}{\mbox{$G_{\mbox{\rm {\scriptsize Avoid-Ramsey}}}$}}
\newcommand{\gmisereramsey}{\mbox{$G_{\mbox{\rm {\scriptsize Avoid'-Ramsey}}}$}}
\newcommand{\gmultiramsey}[1]{\mbox{$G_{\mbox{\rm {\scriptsize Avoid-Ramsey}}^{#1}}$}}
\newcommand{\gachieveramsey}{\mbox{$G_{\mbox{\rm {\scriptsize Achieve-Ramsey}}}$}}
\newcommand{\gachieveramseynt}{\mbox{$G_{\mbox{\rm {\scriptsize Achieve'-Ramsey}}}$}}
\newcommand{\gweakachieveramsey}{\mbox{$G_{\mbox{\rm {\scriptsize Achieve''-Ramsey}}}$}}
\newcommand{\gasymmetricramsey}{\mbox{$G_{\mbox{\rm {\scriptsize Asymmetric-Avoid-Ramsey}}}$}}
\newcommand{\gramseyplus}{\mbox{$G_{\mbox{\rm {\scriptsize Avoid-Ramsey}}^+}$}}
\newcommand{\gpos}{\mbox{$G_{\mbox{\rm {\scriptsize Achieve-POS-CNF}}}$}}
\newcommand{\gposdnf}{\mbox{$G_{\mbox{\rm {\scriptsize Achieve-POS-DNF}}}$}}
\newcommand{\tri}[3]{\triangle(#1\setcomma #2\setcomma #3)}
\newcommand{\trifi}[3]{\{#1\setcomma #2\},\{#1\setcomma #3\},\{#2\setcomma #3\}}
\newcommand{\trif}[3]{\left\{\rule[-1ex]{0cm}{2ex}\trifi{#1}{#2}{#3}\right\}}
\newcommand{\vrii}[3]{\{#1\setcomma #2\},\{#1\setcomma #3\}}
\newcommand{\vri}[3]{\left\{\rule[-1ex]{0cm}{2ex}\vrii{#1}{#2}{#3}\right\}}
\newcommand{\arrowing}{{\sc Arrowing}}
\newcommand{\QBF}{{\sc Quantified Boolean Formula}}

\title{Graph Ramsey games}

\author{Wolfgang Slany\affiliation{mailto:\,wsi@dbai.tuwien.ac.at, 
http://www.dbai.tuwien.ac.at/staff/slany/}}

\begin{document}
\maketitle

\rem{\vfill
\begin{abstract}
\vfill
\noindent We consider combinatorial avoidance and achievement games
based on graph Ramsey theory: The players take turns in coloring still
uncolored edges of a graph $G$, each player being assigned a distinct
color, choosing one edge per move. In avoidance games, completing a
monochromatic subgraph isomorphic to another graph $A$ leads to
immediate defeat or is forbidden and the first player that cannot move
loses. In the avoidance$^+$ variants, both players are free to choose
more than one edge per move. In achievement games, the first player
that completes a monochromatic subgraph isomorphic to $A$ wins.
\citet{Erdos73} were the first to identify some tractable subcases of
these games, followed by a large number of further studies. We
complete these investigations by settling the complexity of all
unrestricted cases: We prove that general graph Ramsey avoidance,
avoidance$^+$, and achievement games and several variants thereof are
\pspacec. We ultra-strongly solve some nontrivial instances of graph
Ramsey avoidance games that are based on symmetric binary Ramsey
numbers and provide strong evidence that all other cases based on
symmetric binary Ramsey numbers are effectively intractable.

\vfill
\end{abstract}
\vfill

\newpage 
}

\tableofcontents

\section{Introduction and overview}

To illustrate the nature of combinatorics, \citet{Cameron94} uses the
following simple game: Two players, Red and Green, compete on a game
board composed of six vertices and all ${6\choose2}=15$ possible edges
between these vertices. The players alternate in coloring at each move
one so far uncolored edge using their color, with the restriction that
building a complete subgraph with three vertices whose edges all have
the same color (a monochromatic triangle) is forbidden. The game ends
when one player is forced to give up because there are no legal moves
left or when one of the players builds a triangle by mistake.

This game was first described under the name Sim by \citet{Simmons69}
in 1969. Since then, it has attracted much interest
\cite{Beck82,Beck81,Beck83,Beck97,Cameron94,Cook79,Cornelius91,Darby98,%
DeLoach71,Engel72,Erdos73,Exoo80,Funkenbusch71,Gardner73a,Gardner73b,%
Gardner86,Hajnal84,Harary82,Knor96,Komjath84,Mead74,Moore87,Nairn73,%
OBrian78,Pekec96,Richer99,Rounds74,Schwartz79,Schwartz81,Shader78,Shader80,Slany88,Zhu97}.
Figure~\ref{fig:sim}
\begin{figure}
\epsfig{file=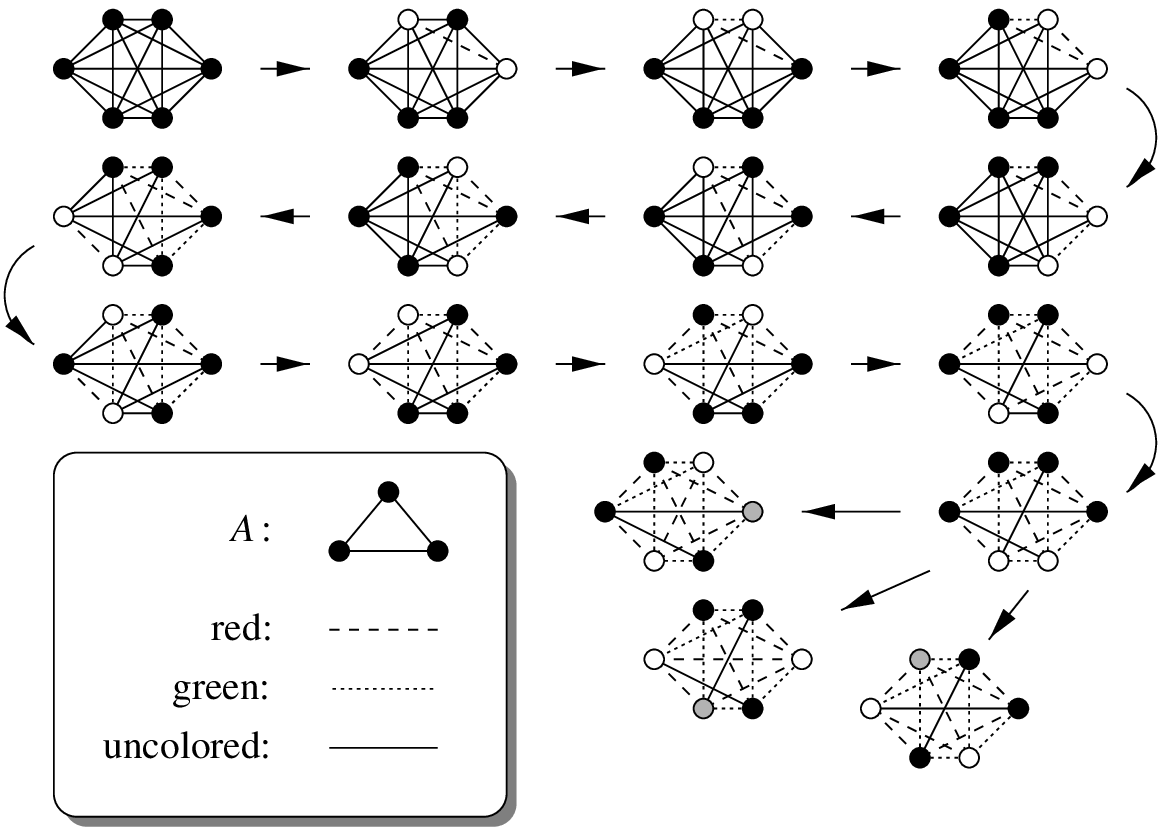,width=\textwidth} 
\boxcaption{Sample play sequence of Sim. The initial, uncolored
game board is shown on the top left corner. Player Red (= dashed
lines) starts by coloring some edge, then player Green (= dotted
lines) colors another one, etc. Finally, Red is forced to give up
since any further coloring would complete a red triangle (= a
monochromatic subgraph isomorphic to $A$).}{fig:sim}
\end{figure}
shows a typical play sequence\footnote{Considering that a hands-on
session with an interactive system often is worth more than a thousand
images, you might want to challenge a Java applet at
http://www.dbai.tuwien.ac.at/proj/ramsey/ that plays Sim and its
avoidance$^+$ variant Sim$^+$, playing perfectly when possible and
improving its strategy by playing over the Internet when perfect play
is impossible. In case you win, you will be allowed to leave your name
in our hall-of-fame!\label{foot:java}}.

Besides their value as motivational examples for the field of
combinatorics, games such as Sim are of practical interest because
they can serve as models that simplify the analysis of competitive
situations with opposing parties that pursue different interests, or
for situations where one is faced with an unforeseeable environment
such as Nature. It is easy to see that playing against a perfectly
intelligent opponent with unlimited computational resources is the
worst case that can happen. If the problem of winning against such an
opponent can be solved, one will also be able to handle all other
eventualities that could arise. Finding a winning strategy for a
combinatorial game can thus be translated into finding a strategy to
cope with many kinds of real world problems such as found in
telecommunications, circuit design, scheduling as well as a large
number of other problems of industrial relevance
\cite{Fraenkel94,Graham90,Papadimitriou85}. Proving or at least
providing strong evidence that finding such a winning strategy is of
high complexity helps to explain the great difficulties one often
faces in corresponding real world problems
\cite{Fraenkel91,Garey79,Papadimitriou94}.  As usual, we mean the
complexity of deciding whether the first player has a winning strategy
when we speak of the complexity of a game in the rest of this paper.

Another, more psychological reason why humans may be attracted by
combinatorial games such as Sim is that they appeal to
\begin{quote}
our primal beastly instincts; the desire to corner, torture, or at
least dominate our peers. An intellectually refined version of these
dark desires, well hidden under the fa\c{c}ade of scientific research,
is the consuming strive ``to beat them all'', to be more clever than
the most clever, in short --- to create the tools to {\em
Math-master\/} them all in hot {\em comb\/}inatorial {\em comb\/}at!
(\citet{Fraenkel94})
\end{quote}

\noindent In Section~\ref{sec:preliminaries}, we define the necessary
notions from combinatorial games, computational complexity and Ramsey
theory, informally introduce graph Ramsey games and discuss previous
work that includes some tractable subcases.  The exact definitions of
all games we study are given in Section~\ref{sec:defs}.
Section~\ref{sec:results} contains our main complexity results on the
previously defined games. Section~\ref{sec:proofs} contains the
detailed proofs for all our complexity results. These results imply
that the unrestricted graph Ramsey games are at least as hard as a
large number of well-known games (e.g., Go~\cite{Lichtenstein80}) and
problems of industrial relevance (e.g., decision-making under
uncertainty such as stochastic scheduling~\cite{Papadimitriou85})
generally recognized as very difficult. Section~\ref{sec:furtherres}
contains complexity results on some further variants of graph Ramsey
games. In Section~\ref{sec:concrete}, we turn to concrete game
instances and present our implemented winning strategy for Sim. We
sketch the heuristics our program uses when perfect play is not
possible, present a winning strategy for the avoidance$^+$ variant
Sim$^+$ of Sim, and provide strong evidence that graph Ramsey
avoidance games based on symmetric binary Ramsey numbers greater for
$n>3$ are intractable from all practical points of view. In
Section~\ref{sec:open}, we state a number of conjectures and open
problems related to Ramsey games.\rem{overview ok?}

\section{Preliminaries and related work}\label{sec:preliminaries}

Like many other combinatorial games, including Chess, Checkers, and
Go, Sim is a two-player zero-sum perfect-information (no hidden
information as in some card games, so there is no bluffing) game
without chance moves (no rolling of dice). Zero-sum here means that
the outcome of the game for the two players is restricted to either
win-loss, loss-win, tie-tie, or draw-draw. The distinction between a
draw and a tie is that a tie ends the game, whereas in a draw, the
game would continue forever, both players being unable to force a win,
following the terminology in the survey on combinatorial games by
\citet{Fraenkel94}. Sim is based on the simplest nontrivial example
of Ramsey theory~\cite{Graham90,Nesetril95}, the example being also
known under the name of ``party-puzzle'': How many persons must be at
a party so that either three mutual acquaintances or three persons
that are not mutual acquaintances are present? More formally, classic
binary Ramsey numbers are defined as follows:

\begin{defn}{$\ramsey(n,m)$} \label{def:classic-ramsey-nr}
denotes the smallest number $r$ such that any complete graph $K_r$ (an
undirected graph with r vertices and all possible edges between them)
whose edges are all colored in red or in green either contains a red
subgraph isomorphic to $K_n$ or a green subgraph isomorphic to $K_m$.
In classic\/ {\em symmetric} binary Ramsey numbers, $n$ equals $m$.
\end{defn}

\begin{obs}
Another equivalent formulation says that $\ramsey(n,m)$ is the
smallest number of vertices such that an arbitrary undirected graph of
that size either contains an $n$-clique (that is, a $K_n$) or an
$m$-independent set ($m$ isolated vertices, that is, with no edges
between them).
\end{obs}

\noindent The classic result of F.\ P.\ \citet{Ramsey30}, a structural
generalization of the pigeon-hole principle, tells us that these
numbers always exist:

\begin{theo}[\citet{Ramsey30}]\label{theo:ramsey} $\forall (n,m)\in\N^2 \quad\ramsey(n,m)<\infty.$
\end{theo}

\noindent Ramsey used this result (that by itself was popularized only
some years later by \citet{Erdos35}) to prove that if $\phi$ is a
first-order formula of the form $\exists x_1 \exists x_2 \cdots
\exists x_n \forall y_1 \cdots \forall y_m \Phi$ where $\Phi$ is
quantifier free, i.e., if $\phi$ is a Bernays-Sch\"{o}nfinkel formula,
then the problem whether $\phi$ holds for every finite structure is
decidable (see also \citet{Nesetril95}).

A simple combinatorial argument that $\ramsey(3,3)=6$ is shown in
Figure~\ref{fig:r33e6}, and so the minimal number of persons
satisfying above's ``party-puzzle'' question is six.
\begin{figure}
\begin{center}\epsfig{file=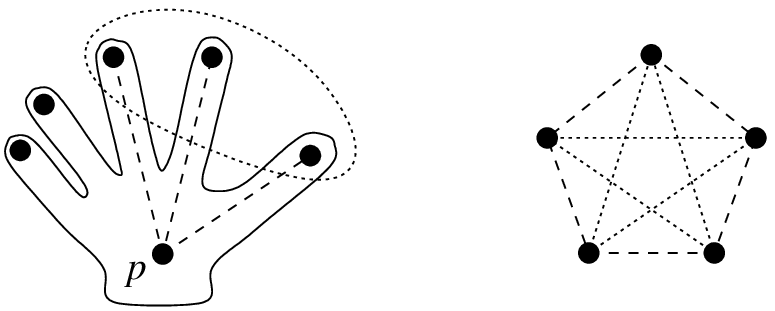,width=.75\textwidth}\end{center}
\boxcaption{Visual proof that $\ramsey(3,3)=6$, as communicated by
Ranan Banerji. The drawing on the left shows that six vertices are
enough, as follows: Take any vertex $p$ (as in `palm') of an
edge-2-colored $K_6$.  At least three edges connected to $p$ will have
the same color. Without loss of generality, assume that this color is
the dashed one. Consider the three vertices connected to $p$ through
these three edges: Either one of the edges that connect two of these
vertices is of the dashed type (and then there is a dashed triangle
with the edges connected to $p$), or not (and then the three top edges
form a triangle in the other color). The edge-2-colored $K_5$ on the
right serves as a counter-example, showing that five vertices are
insufficient to force a monochromatic triangle. Thus, six is the
smallest number with the required property.}{fig:r33e6}
\end{figure}
Theoretically, Sim ends after a maximum of 15 moves since this is the
number of edges in a complete graph with six vertices. If we define
Sim such that monochromatic triangles are not allowed, and since
Ramsey theory says that any edge-2-colored $K_6$ will contain at least
one monochromatic triangle, we know that the second player will not be
forced to give up simply because all edges are colored after 15 moves,
as all games will end before the 15$^{th}$ move. The game of Sim as it
is usually described and played ends when one of the players completes
a triangle in his color, whether forced or by mistake (this is called
a `mis\`ere-type' end condition: the last player to {\em move\/}
loses, see e.g.\ \citet{Guy91}), with no winner, that is, a tie,
defined for the case when all edges are colored without a
monochromatic triangle having been completed. For this
mis\`ere-variant of Sim, the $\ramsey(3,3)=6$ result implies that no
game will ever end in a tie.

It is easy to see that in finite, two-player zero-sum
perfect-information games with no ties and no chance moves, either the
player who starts the game or his opponent must have the possibility
to play according to a {\em winning strategy}: A player who follows
such a strategy will always win no matter how well the opponent plays
(for the existence of such a strategy, see for instance the
fundamental theorem of combinatorial game theory in
\citet{Fraenkel91}). Clearly, this means that one of the players will
have an a-priori upper-hand in Sim, so the answer to the following
question is of central interest: Which of the two players has a
winning strategy, the first or the second to move?

This question is the classic decision problem one can ask for any
combinatorial game. In case of Sim, \citet{Mead74} have shown that the
second player can always win. Nevertheless, a winning strategy that is
easy to memorize for human players has so far eluded us, despite much
effort \cite{Cook79,DeLoach71,%
Funkenbusch71,Gardner73a,Gardner73b,Gardner86,Mead74,Moore87,Nairn73,%
OBrian78,Rounds74,Schwartz79,Schwartz81,Shader78,Shader80,Slany88,Zhu97}.
Knowing the strategy itself, especially if it can be stated in a
concise form, might appear to be even better, but it is easy to see
that knowing the answer to the decision problem for an arbitrary game
situation, or at least being able to efficiently find out that answer,
is equivalent to knowing the complete strategy.

A game being finite means that it should theoretically be possible to
solve it. However, the trouble is that it might take an astronomical
amount of time and memory (often even more as we will see in
Section~\ref{sec:sim4}) to actually compute the winning strategy. Note
that J.~\citet{Schaeffer96a} have started trying to prove that a
certain strategy for the game of Checkers is a winning one, using a
massive amount of parallel hardware already running for several
years. Their attempt requires the analysis of positions roughly equal
in number to the square root of the size of the full game tree of
Checkers (which in case of Checkers appears to be barely in reach of
present day computing power) and thus can be substantially faster than
finding a winning strategy from scratch, obviously for the price that
in case their strategy is shown not to be a winning one, the game
remains unsolved. Being usually unable to even prove that a strategy
is a winning one, we turn to the next best thing, which is to classify
the games in terms of computational complexity classes, that is, to
find out how the function bounding the computational resources that
are needed in the worst case to determine a winning strategy for the
first player grows in relation to the size of the game description.

Here a technical problem becomes apparent, in that games must be
scalable instead of having a fixed finite size in order to be
classifiable. Generalizations to boards of size $n\times n$ of
well-known games such as Chess, Checkers, and Go have been classified
as \pspacec\ and \exptimec\
\cite{Fraenkel78,Fraenkel81,Lichtenstein80}. \pspace\ in particular is
important for the analysis of these and large classes of more formal
combinatorial games
\cite{Fraenkel91,Fraenkel94,Garey79,Papadimitriou94,Schaefer78}. \pspace\
is the class of problems that can be solved using memory space bounded
by a polynomial in the size of the problem description. \pspacec\
problems are the hardest problems in the class \pspace: Solving one of
these problems efficiently would mean that we could solve {\em any\/}
other problem in \pspace\ efficiently as well. While nobody so far was
able to show that \pspace\ problems are inherently difficult, despite
much effort to show that the complexity class \p\ containing the
tractable problems solvable in polynomial time is different from
\pspace, it would be very surprising if they were not. Indeed, the
well-known complexity class \np\ is included in \pspace, so problems
in \pspace\ are at least as difficult as many problems believed to be
very hard such as the satisfiability of boolean formulas or the
traveling salesman problem. This means that it is rather unlikely that
efficient general algorithms to solve \pspacec\ combinatorial games do
exist. For further details on computational complexity theory, consult
\citet{Garey79} or \citet{Papadimitriou94}. Obviously, the high
complexity of such combinatorial games contributes to their
attractiveness.

So the question is, what could be a generalization of Sim to game
boards of arbitrary size?  Let us first introduce some more notions
from graph Ramsey theory that generalize the classic Ramsey numbers from
Definition~\ref{def:classic-ramsey-nr}:

\begin{defn}[see, e.g., \cite{Burr87,Diestel97,Graham90,Schaefer99}]\label{def:arrowing}
{$G\rightarrow (A^r,A^g)$:} We say that a graph $G$\/ {\em arrows} a
graph-tuple $(A^r,A^g)$ if for every edge-coloring with colors red and
green, a red $A^r$ or a green $A^g$ occurs as a subgraph. In\/ {\em
symmetric} arrowing, $A^r=A^g=A$, and $G$ is called a\/ {\em Ramsey
graph} of $A$ if $G\rightarrow A$.
\end{defn}

\begin{obs}
$K_{\scriptsize \ramsey(n,m)}\rightarrow (K_n,K_m).$
\end{obs}

\noindent The following generalization of Theorem~\ref{theo:ramsey}
was proved in 1962 by \citet{Harary62} after hearing a lecture on
Ramsey theory given by Erd\H{o}s and first published around 1973,
independently by \citet{Chvatal72}, by \citet{Deuber75}, by
\citet{Erdos75}, and by \citet{Roedl73}:

\begin{theo}[see, e.g., \cite{Diestel97}]
Every graph has Ramsey graphs. In other words, for every graph $A$
there exists a graph $G$ that, for every edge-coloring with colors red
and green, either contains a red or a green subgraph isomorphic to
$A$.
\end{theo}

\noindent Note that the complexity of the
arrowing relation has recently been determined:

\begin{defn}{\arrowing}\\[1ex]
Instance: (Finite) graphs $G$, $A^r$, and $A^g$.\\
Question: Does\/ $G\rightarrow (A^r,A^g)$ ?
\end{defn}

\begin{theo}[M.~\citet{Schaefer99}]\label{theo:piptc}\quad
\arrowing\ is\/ \piptc.
\end{theo}

\noindent We extend the arrowing relation for our purposes:

\begin{defn}{$(G,E^r,E^g)\rightarrow A$:}\label{def:precolarr}
A\/ {\em partly} edge-colored graph $(G,E^r,E^g)$, where some edges
$E^r$ of $G$ are precolored in red and some other edges $E^g$ of $G$
are precolored in green, arrows a graph $A$ if every\/ {\em
complete} edge-coloring of $(G,E^r,E^g)$ with colors red and green
contains a monochromatic subgraph isomorphic to $A$.
\end{defn}

\noindent The game \gramsey\ is the generalization of Sim to graph
Ramsey theory (exact definitions of all graph Ramsey game variants
follow in Section~\ref{sec:defs}). Similarly, \gachieveramsey\ is a
graph Ramsey achievement game.  \citet{Harary82} studied both
\gachieveramsey\ and \gramsey\ where $G$ is restricted to complete
graphs, $A$ being an arbitrary graph. We call \gramseyplus\ the
avoidance$^+$ variant where each player selects {\em at least one\/}
so-far uncolored edge per move. For graph Ramsey achievement games,
several tractable subcases are known:

\begin{theo}[\citet{Erdos73}]\label{theo:lseekramsey} The first player has a winning strategy in 
\gachieveramsey$(K_n,K_k,\{\},\{\})$ if
\[k\;\leq\;\frac{1}{2}\;\log_2 n\]
and the game ends in a\/ {\em tie} if
\[2^{\,l}\;>\;{n\choose k}, \quad\mbox{ where }\quad l\;=\;{{k\choose 2}\;-\;1}\;,\]
i.e., it is a\/ {\em tie} if 
\[k\;\geq\;2\;(1\;+\;o(1))\;\log_2 n\;.\] 
\end{theo}

\noindent While these results do not cover all cases with complete
graphs such as for example
Sim$_a\;=\;\gachieveramsey(K_6,K_3,\{\},\{\})$, small instances of
\gachieveramsey\ generally seem to be very easy to analyze.
\begin{figure}
\epsfig{file=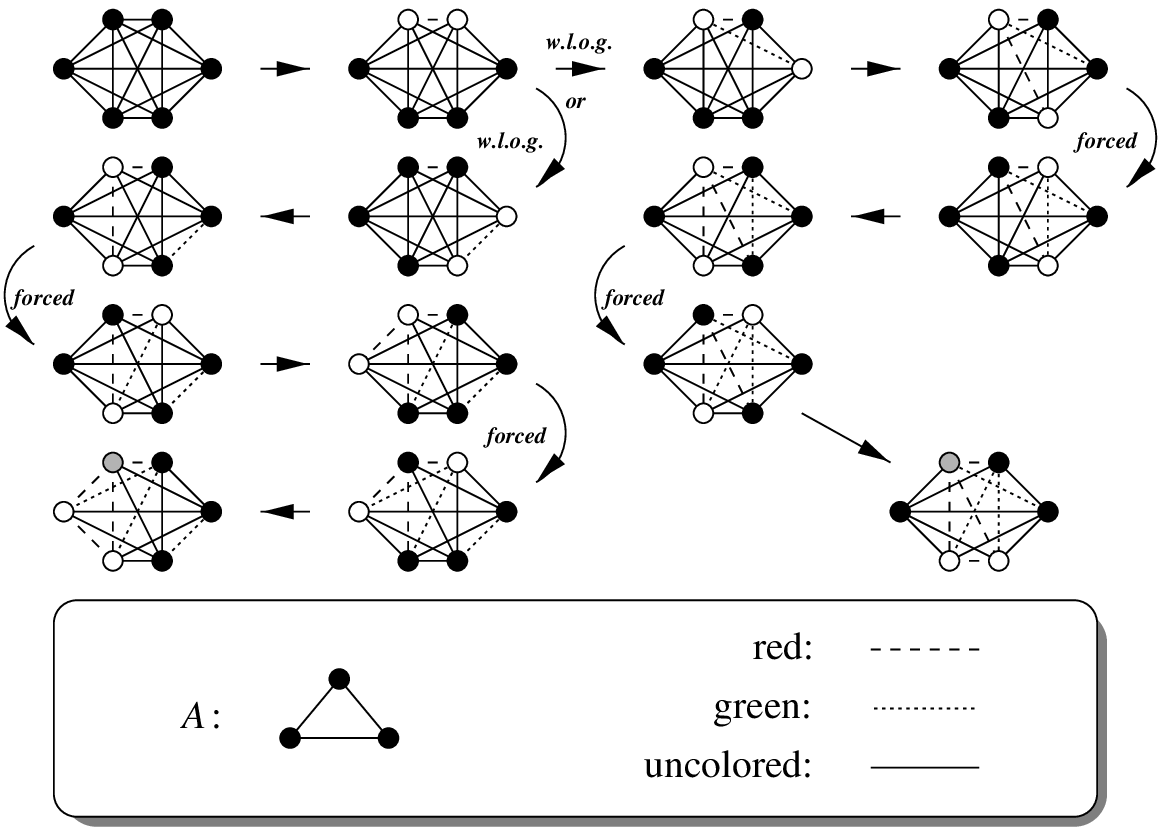,width=\textwidth} \boxcaption{Trivial winning
strategy for player Red in
Sim$_a\;=\;\gachieveramsey(K_6,K_3,\{\},\{\})$.}{fig:achieveramsey}
\end{figure}
Figure~\ref{fig:achieveramsey} shows, for instance, a trivial winning
strategy for the first player in Sim$_a$. \citet{Beck81} and
\citet{Beck82} have generalized these results to games where the
players alternate in choosing among previously unchosen elements of
the complete $k$-uniform hypergraph of $N$ vertices $K^k_N$, and the
first player wins if he has selected all $k$-tuples of an $n$-set. For
the case $k=2$, their results subsumes
Theorem~\ref{theo:lseekramsey}. They also study infinite Ramsey games
where the edges of the hypergraphs are required to be infinite but
countable, for which they show that there always exist winning
strategies for the first player. Several games of this kind are
analyzed, all featuring simple winning strategies that imply their
tractability. Further studies following the results of \citet{Erdos73}
can be found in
\cite{Beck83,Beck97,Exoo80,Hajnal84,Knor96,Komjath84,Pekec96,Richer99}.

\section{Definitions of the graph Ramsey games}\label{sec:defs}

Let us now precisely define the introduced graph Ramsey games. Note
that in the following definitions, the precolorings $(E^r,E^g)$ are
part of the input and are needed for the analysis of arbitrary game
situations appearing in mid-game.

\begin{defn}\label{def:gramsey} \ The graph Ramsey\/ {\em avoidance} game 
\gramsey$(G,A,E^r,E^g)$ is played on a graph $G=(V,E)$, another graph
$A$, and two nonintersecting sets $E^r \cup E^g \subseteq E$ that
contain edges initially colored in red and green, respectively. Two
players, Red and Green, take turns in selecting at each move one
so-far uncolored edge from $E$ and color it in red for player Red
respectively in green for player Green. However, both players are
forbidden to choose an edge such that $A$ becomes isomorphic to a
subgraph of the red or the green part of $G$. It is Red's turn. The
first player unable to move loses.
\end{defn}

\begin{defn}\label{def:gmisereramsey} 
\gmisereramsey\ is the mis\`ere-variant of \gramsey, were completing a
monochromatic subgraph isomorphic to graph $A$ leads to immediate
defeat. The game ends in a tie if no edges are left to color.
\end{defn}

\noindent Clearly, these two avoidance variants coincide whenever
$(G,E^r,E^g)\rightarrow A$ (the proof is straightforward):

\begin{corol}\label{cor:misere}
If $(G,E^r,E^g)\rightarrow A$, then player Red has a winning strategy
in \gramsey$(G,A,E^r,E^g)$ iff player Red has a winning strategy in
\gmisereramsey$(G,A,E^r,E^g)$.
\end{corol}

\begin{obs} \ Sim $=$ \gramsey$(K_{\scriptsize
\ramsey(3,3)},K_3,\{\},\{\})$\\
\hspace*{38mm} $=$ \gmisereramsey$(K_{\scriptsize
\ramsey(3,3)},K_3,\{\},\{\})$.
\end{obs}

\noindent The following avoidance$^+$ variant intuitively corresponds
even closer to the spirit of Ramsey theory because any combination in
the number of red and green edges is possible (in the other graph
Ramsey avoidance and achievement games, red and green edges are added
at the same rate):

\begin{defn}{\gramseyplus$(G,A,E^r,E^g)$:}\label{def:gramseyplus}
Everything is as in Definition~\ref{def:gramsey}, except that each
player selects\/ {\em at least one} so-far uncolored edge from $E$
during one move.
\end{defn}

\begin{obs} \
Sim$^+$ $=$ \gramseyplus$(K_{\scriptsize \ramsey(3,3)},K_3,\{\},\{\})$.
\end{obs}

\noindent In the case of graph Ramsey achievement games, three major
variants can be distinguished, as follows:

\begin{defn} \ In the graph Ramsey\/ {\em achievement} game 
\gachieveramsey$(G,A,E^r,E^g)$ everything is as in
Definition~\ref{def:gramsey}, except that the first player who builds
a monochromatic subgraph isomorphic to $A$\/ {\em wins}.
\end{defn}

\begin{defn} A simple strategy-stealing argument tells us that with
optimal play on an uncolored board, \gachieveramsey\ must be either a
first-player win or a draw, so it is only fair to count a draw as a
second-player win. Let us call this variant \gachieveramseynt.
\end{defn}

\noindent We know from the fundamental theorem of combinatorial game
theory (see e.g.\ \cite{Fraenkel91}) that there exists a winning
strategy for this game. It is straightforward that when
$(G,E^r,E^g)\rightarrow A$, \gachieveramsey\ and \gachieveramseynt\
are in fact the same game.

\begin{defn} 
Following the terminology of \citet{Beck82}, let us call the variant
of\/ \gachieveramsey\ where all the second player does is to try to
prevent the first player to build $A$, without winning by building it
himself, the ``weak'' graph Ramsey achievement game
\gweakachieveramsey.
\end{defn}

\noindent Again, it is straightforward that when the first player has
a winning strategy or when there is no possibility for the second
player to build a green subgraph isomorphic to $A$, \gachieveramseynt\
and \gweakachieveramsey\ are in fact the same game.

\begin{obs} \
Sim$_a$ $=$ \gachieveramsey$(K_{\scriptsize
\ramsey(3,3)},K_3,\{\},\{\})$,\\
\qquad and from the point of view of a perfect first player,\\
\hspace*{30.8mm} Sim$_a$ $=$ \gachieveramseynt$(K_{\scriptsize
\ramsey(3,3)},K_3,\{\},\{\})$\\
\hspace*{39.7mm} $=$ \gweakachieveramsey$(K_{\scriptsize \ramsey(3,3)},K_3,\{\},\{\})$.
\end{obs}

\section{Main complexity results}\label{sec:results}

We have long believed that the problems of deciding whether the first
players have winning strategies in the graph Ramsey avoidance games
are complete for polynomial space. We show here that our intuition was
indeed right, corroborating the apparent difficulty of Sim and its
\gramseyplus\ variant Sim$^+$. All game variants mentioned below have
been formally defined in Section~\ref{sec:defs}. The proofs of these
results are discussed in Section~\ref{sec:proofs}.

\begin{theo}\label{theo:gramsey} \ \gramsey\ is\/ \pspacec.
\end{theo}

\begin{theo}\label{theo:gmisereramsey} \ \gmisereramsey\ is\/ \pspacec.
\end{theo}

\noindent In order to prove these results, a special gadget
construction was needed that constrains the moves of the players in
spite of their apparent freedom to choose any uncolored edge. Its
construction was inspired from the similar notion of ``illegitimate''
moves introduced by \citet{Even76} and further developed by
T.~\citet{Schaefer78}.

\begin{theo}\label{theo:gramseyplus} \ \gramseyplus\ is\/ \pspacec.
\end{theo}

\noindent Theorem~\ref{theo:gramseyplus} facilitates the matching
between abstract problems and real life applications as it allows to
drop the artificial requirement that players must move in a
predetermined sequence. Let us observe, however, that \pspacec{}ness
of avoidance games such as the avoidance games played on propositional
formulas and on sets described in \cite{Schaefer78} do not
automatically imply the \pspacec{}ness of their avoidance$^+$
variants: Most of these \pspacec\ single-choice-per-move avoidance
games have trivially decidable, and thus tractable, avoidance$^+$
variants.  We also note that even in case both the avoidance$^+$ and
the single-choice-per-move avoidance variant are \pspacec, it is easy
to see that the players having winning strategies can be different for
the two games, and that even if in both games the first player has a
winning strategy, completely new game situations requiring different
playing behavior may arise in an avoidance$^+$ variant.

\begin{obs} Note that a mis\`ere-variant of\/ \gramseyplus\ is easily 
imaginable, its\/ \pspacec{}ness proof following the lines of the proof
of Theorem~\ref{theo:gmisereramsey} when applied to
Theorem~\ref{theo:gramseyplus}.
\end{obs}

\begin{corol}\label{cor:avoidrestriction} \
\gramsey, \gmisereramsey, and\/ \gramseyplus\ remain\/ \pspacec\ even
if the avoidance graph $A$ is restricted to a specific fixed graph.
\end{corol}

\noindent For achievement games, the situation is similar:

\begin{theo}\label{theo:gweakachieveramsey} \ 
\gweakachieveramsey\ is\/ \pspacec. 
\end{theo}

\begin{theo}\label{theo:gachieveramseynt} \ 
\gachieveramseynt\ is\/ \pspacec.
\end{theo}

\begin{corol}\label{cor:achieverestriction} \
\gweakachieveramsey\ and\/ \gachieveramseynt\ remain\/ \pspacec\ even if
the achievement graph $A$ is degree-restricted.
\end{corol}

\begin{theo}\label{theo:gachieveramsey} \ 
\gachieveramsey\ is\/ \pspacec.
\end{theo}

\noindent The \pspacec{}ness of the achievement games came a bit as a
surprise since tractable subcases are
known~\cite{Beck81,Beck83,Erdos73,Exoo80,Hajnal84,Harary82,Knor96,Komjath84,Pekec96},
and the \gachieveramsey\ game Sim$_a$ corresponding to Sim with
respect to graphs $G$ and $A$ has a trivial winning strategy, in
blatant contrast to Sim and Sim$^+$.

\section{Proofs of the main complexity results}\label{sec:proofs}

One way to prove \pspacec{}ness consists in showing that the problem
is solvable in \pspace\ (``membership'') and that it is at least as
difficult as any other problem in \pspace\ (``hardness''). The
intricate parts of the proofs of
Theorems~\ref{theo:gramsey}--\ref{theo:gachieveramsey} will be found
in their hardness parts. The following lemma establishes the
membership parts of all proofs:

\begin{lem}\label{lemma:membership}
All graph Ramsey games defined in Section~\ref{sec:defs} are in
\pspace.
\end{lem}

\proof\ Let $n\de |(G,A,E^r,E^g)|$ denote the size of the input. The
number of moves in any graph Ramsey game is limited by the number of
initially uncolored edges in the graph $G$, so any game will end after
at most $|E|-|E^r|-|E^g|<n$ edge colorings, and each game situation
can be described as the edge that was just colored, so this
information uses memory $O(\log n)$ which is bounded by $O(n)$. It is
easy to enumerate in some lexicographic order all game situations that
can originate from a particular game situation through the coloring of
one edge. Altogether, this implies membership in
\pspace\ by the following argument: Given an initial game situation, a
depth-first algorithm that checks all possible game sequences but
keeps in memory only one branch of the game tree at a time,
backtracking to unexplored branching points in order to scan through
the whole game tree, can decide whether there is a winning strategy
for player Red using memory bounded by the maximum stack size, which
is $\,O(n \log n)\:<\:O(n^2)$ and thus polynomial in the size of the
input. \halmoseol

\sproof{theo:gramsey}

Membership of \gramsey\ in \pspace\ follows from
Lemma~\ref{lemma:membership}. To show hardness, i.e., that a problem
is at least as difficult as {\em any\/} other problem in the class, it
is enough to show that it is at least as difficult as {\em one\/}
complete problem from that class. Thus, it suffices to show that there
exists a simple reduction from one known \pspacec\ problem to
\gramsey. In the case at hand, the complete problem will be \gpos, a
game first described by T.~\citet{Schaefer78}. The definition of the
game \gpos\ is restated below in Definition~\ref{def:gpos}. The
reduction will be a \logspace\ transducer that transforms any instance
of the \gpos\ game into an instance of the \gramsey\ game using only
space logarithmic in the size of the \gpos\ instance for intermediate
results, and such that the answer to the \gpos\ decision problem is
the same as the answer to the corresponding \gramsey\ decision
problem. This would allow to decide \gpos, which is know to be
complete and therefore by definition a most difficult problem in
\pspace, by doing a simple transformation and solving a \gramsey\
problem, thus establishing that deciding \gramsey\ has to be at
least as difficult as the complete problem of deciding \gpos, and
therefore as difficult as {\em any\/} other problem in \pspace.

The game that is reduced to \gpos\ is defined as follows:

\begin{defn}[T.~\citet{Schaefer78}]\label{def:gpos} \gpos$(F)$: We 
are given a positive CNF formula $F$. A move consists of choosing some
variable of $F$ which has not yet been chosen. Player I starts the
game. The game ends after all variables of $F$ have been
chosen. Player I wins iff $F$ is\/ {\tt true} when all variables chosen
by player I are set to\/ {\tt true} and all variables chosen by player
II are set to\/ {\tt false}. \end{defn}

\begin{obs}\label{obs:gposwbg}
\gpos\ by definition is a finite two-player zero-sum
perfect-information game with no ties and no chance moves, so either
one of its two players has a winning strategy.
\end{obs}

\noindent For example, on input $x_1\And (x_2 \Or x_3)\And (x_2 \Or
x_4)$ player II has a winning strategy, whereas on input $(x_1\Or
x_4)\And (x_2\Or x_3)\And (x_2\Or x_4)$ player I has a winning
strategy.

To prove the hardness part of the proof of Theorem~\ref{theo:gramsey},
formally we will show that
\[\gpos\,\le_{\mbox{log}}\,\gramsey.\] The following result from
T.~\citeauthor{Schaefer78} would then complete our proof:

\begin{theo}[T.~\citet{Schaefer78}] ~\\
\hspace*{\parindent}\gpos\ is \pspacec.
\end{theo}

\begin{figure}[!t]
\begin{center}\epsfig{file=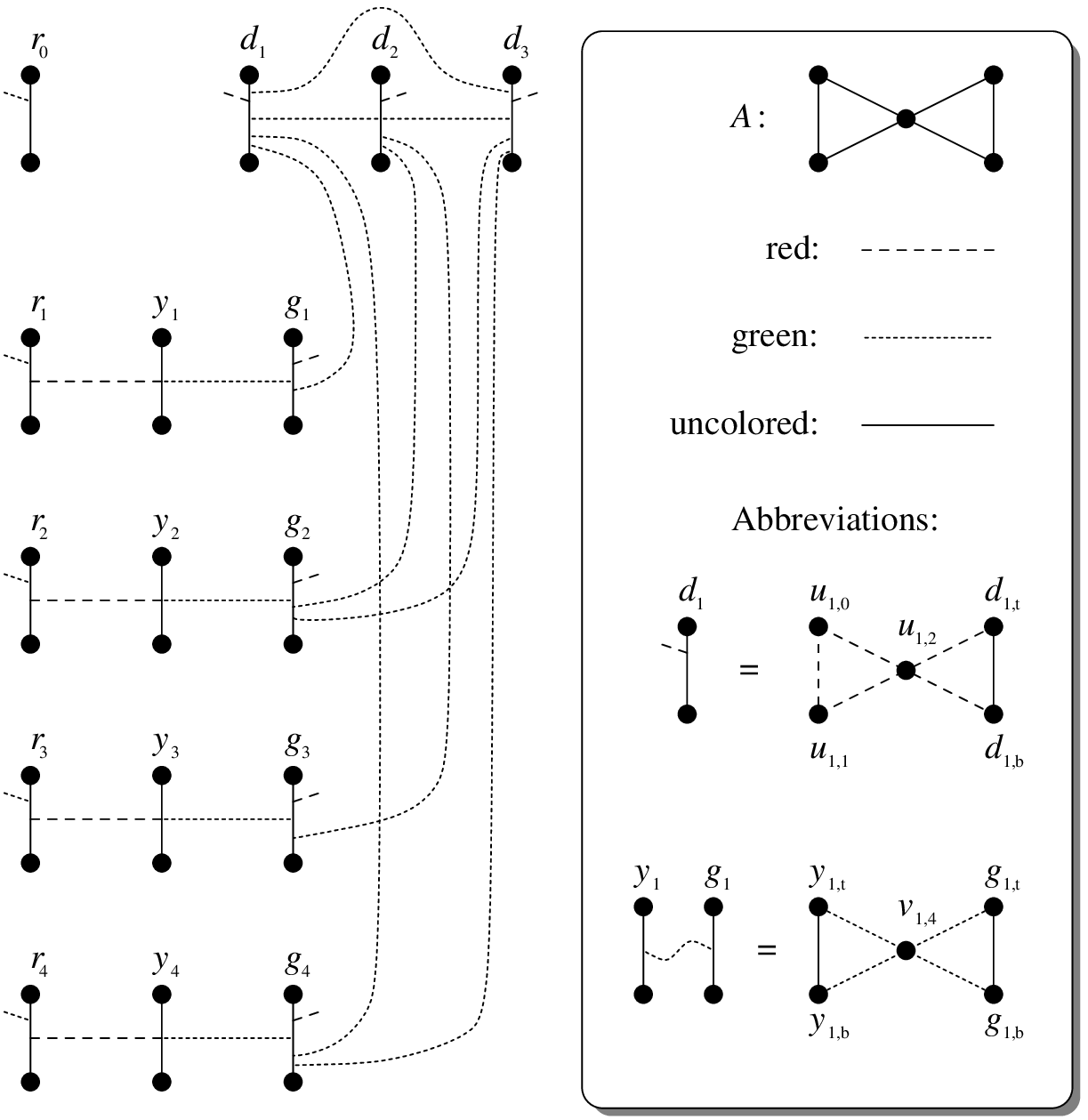,width=.97\textwidth} 
\end{center}
\boxcaption{Example of the reduction
$\gpos\,\le_{\mbox{log}}\,\gramsey$ from the proof of
Theorem~\ref{theo:gramsey}. The graph $G$ is shown on the left and
corresponds to the input formula $F=(x_1\Or x_4)\And (x_2\Or x_3)\And
(x_2\Or x_4)$, featuring a winning strategy for player I. The graph
$A$ (a `bow-tie') that both players must avoid in their color is shown
on the top right corner. The dashed (= red) and dotted (= green) lines
are abbreviations as partly indicated on the bottom right corner, the
rest following the same ideas.}{fig:gramsey}
\end{figure}
\noindent Let us sketch here the idea of the proof using the small example in 
Figure~\ref{fig:gramsey}. The exact description of the reduction
follows later. Each gadget $P_i$, containing among other precolored
edges (note the abbreviations in Figure~\ref{fig:gramsey}) the three
uncolored edges $r_i\setcomma y_i\setcomma g_i$, corresponds to the
boolean variable $x_i$ of $F$. Each gadget $D_j$, containing the
uncolored edge $d_j$, corresponds to conjunct $C_j$ of $F$. The links
between the two types of gadgets correspond to the occurrence of the
variables in the conjuncts. Player Red can only color edges $r_i$ and
$y_i$ whereas player Green can only color edges $g_i$, $y_i$, and
possibly one of the edges $d_j$ if the $g_i$'s connected to it are
uncolored. By counting the number of possible moves, one sees that
Green has a winning strategy if he succeeds in coloring one edge $d_j$
at move $2n+2$. Coloring edge $y_i$ in a $P_i$ gadget means that the
other player can only color the remaining border edge. Thus, the
players first will race to color all edges $y_i$, since by doing so,
Red could possibly hinder Green from coloring any edge $d_j$ at the
end, whereas Green could possibly leave enough edges $g_i$ uncolored
so that he can color one edge $d_j$ at the end.

\begin{fact}\label{fact:iff}
From the definition of \gpos, we easily see that player I wins iff he
succeeds in playing some variable in each conjunct.  This is mirrored
in \gramsey\ as follows: Player Red can win iff he succeeds in
coloring some edge $y_i$ so that player Green later on can only choose
edges $g_i$ in these particular triples, making it impossible for
Green to color any edge $d_j$ at the end.
\end{fact}

\noindent The rest of the proof consists in an analysis of several
cases showing that there is a winning strategy for player I in
\gpos$(F)$ iff there is a winning strategy for player Red in
\gramsey$(G,A,E^r,E^g)$. The detailed proof follows.

We first define the \logspace\ reduction from \gpos$(F)$ to
\gramsey$(G,A,E^r,E^g)$ that has already been illustrated through a
small example in Figure~\ref{fig:gramsey}: Let a positive CNF formula
$F$ be given. Assume without loss of generality that $F = C_1 \And
\ldots \And C_m$ where each conjunct $C_j$ is a disjunction of $n_j$
positive literals, that is, $C_j = l_{j,1} \Or \ldots \Or l_{j,n_j}$
where $l_{j,k} \in \{x_1, \ldots, x_n\}$ and all $n$ variables appear
at least once in $F$.  We then define the graphs $G\de(V,E)$,
$A\de(V^A,E^A)$ and the edge-sets $E^r$, $E^g$, by
\addtolength{\jot}{1ex} \rem{no splitting of formulas over
page-breaks!}
\begin{eqnarray*}
V   & \de & \bigcup_{0 \leq i \leq n} X_i\:,\\
X_0 & \de & \bigcup_{0 \leq j \leq m} B_j\:,\\
B_0 & \de & \{u_{0,0}\setcomma u_{0,1}\setcomma u_{0,2}\setcomma r_{0,\mathrm{t}}\setcomma r_{0,\mathrm{b}}\}\:,\\
B_j & \de & \{u_{j,0}\setcomma u_{j,1}\setcomma u_{j,2}\setcomma d_{j,\mathrm{t}}\setcomma d_{j,\mathrm{b}}\}\union\\[-1ex] 
       && \bigcup_{1 \leq p < j} \{w_{j,p}\}\union  
          \bigcup_{1 \leq k \leq n_j} \{f_{j,k}\} \quad\mbox{for}\quad 1 \leq j \leq m\:,\\
X_i & \de & \{v_{i,0}\setcomma v_{i,1}\setcomma v_{i,2}\setcomma r_{i,\mathrm{t}}\setcomma r_{i,\mathrm{b}}\setcomma v_{i,3}\setcomma y_{i,\mathrm{t}}\setcomma y_{i,\mathrm{b}}\setcomma\\[-1ex] 
       && \quad\!\!\! v_{i,4}\setcomma g_{i,\mathrm{t}}\setcomma g_{i,\mathrm{b}}\setcomma v_{i,5}\setcomma v_{i,6}\setcomma v_{i,7}\} \quad\mbox{for}\quad 1 \leq i \leq n\:,\\
E   & \de & \bigcup_{0 \leq i \leq n} P_i\:,\\
P_0 & \de & \bigcup_{0 \leq j \leq m} D_j\:,\\
D_0 & \de & \tri{u_{0,0}}{u_{0,1}}{u_{0,2}}\union\tri{u_{0,2}}{r_{0,\mathrm{t}}}{r_{0,\mathrm{b}}}\:,\\[-1ex]
       && \hspace{.25\textwidth}\mbox{where} \quad\tri{\alpha}{\beta}{\gamma} \de \trif{\alpha}{\beta}{\gamma}\:,\\
D_j & \de & \tri{u_{j,0}}{u_{j,1}}{u_{j,2}}\union\tri{u_{j,2}}{d_{j,\mathrm{t}}}{d_{j,\mathrm{b}}}\union\\[-1ex]
       && \bigcup_{1 \leq p < j} \left\{\rule[-1ex]{0cm}{2ex}\vrii{w_{j,p}}{d_{p,\mathrm{t}}}{d_{p,\mathrm{b}}},\vrii{w_{j,p}}{d_{j,\mathrm{t}}}{d_{j,\mathrm{b}}}\right\}\union\\[-1ex]
       && \bigcup_{1 \leq k \leq n_j} \left\{\rule[-1ex]{0cm}{2ex}\vrii{f_{j,k}}{d_{j,\mathrm{t}}}{d_{j,\mathrm{b}}},\vrii{f_{j,k}}{g_{h,\mathrm{t}}}{g_{h,\mathrm{b}}}\:|\: l_{j,k}=x_h\right\}\\[-3ex]
       && \hspace{.65\textwidth}\mbox{for}\quad 1 \leq j \leq m\:,\\
P_i & \de & \tri{v_{i,0}}{v_{i,1}}{v_{i,2}}\union\tri{v_{i,2}}{r_{i,\mathrm{t}}}{r_{i,\mathrm{b}}}\union\tri{v_{i,3}}{r_{i,\mathrm{t}}}{r_{i,\mathrm{b}}}\union\\[-1ex]
       && \tri{v_{i,3}}{y_{i,\mathrm{t}}}{y_{i,\mathrm{b}}}\union\tri{v_{i,4}}{y_{i,\mathrm{t}}}{y_{i,\mathrm{b}}}\union\tri{v_{i,4}}{g_{i,\mathrm{t}}}{g_{i,\mathrm{b}}}\union\\[-1ex]
       && \tri{v_{i,5}}{g_{i,\mathrm{t}}}{g_{i,\mathrm{b}}}\union\tri{v_{i,5}}{v_{i,6}}{v_{i,7}}\qquad\qquad\qquad\mbox{for}\quad 1 \leq i \leq n\:,\\
V^A & \de & \{a_0\setcomma a_1\setcomma a_2\setcomma a_3\setcomma a_4\}\:,\\
E^A & \de & \tri{a_0}{a_1}{a_2}\union \tri{a_2}{a_3}{a_4}\:,\\
E^r & \de & \bigcup_{0 \leq i \leq n} P^r_i\:,\\
P^r_0 & \de & \bigcup_{1 \leq j \leq m} \left(\rule[-1ex]{0cm}{2ex}\tri{u_{j,0}}{u_{j,1}}{u_{j,2}}\union\vri{u_{j,2}}{d_{j,\mathrm{t}}}{d_{j,\mathrm{b}}}\right)\:,\\
\end{eqnarray*}
\begin{eqnarray*}
P^r_i & \de & \left\{\rule[-1ex]{0cm}{2ex}\vrii{v_{i,3}}{r_{i,\mathrm{t}}}{r_{i,\mathrm{b}}},\vrii{v_{i,3}}{y_{i,\mathrm{t}}}{y_{i,\mathrm{b}}},\right.\\[-1ex]
         && \ \:\left.\rule[-1ex]{0cm}{2ex}\vrii{v_{i,5}}{r_{i,\mathrm{t}}}{r_{i,\mathrm{b}}}\right\}\union\tri{v_{i,5}}{v_{i,6}}{v_{i,7}}\qquad\mbox{for}\quad 1 \leq i \leq n\:,\\
E^g & \de & \bigcup_{0 \leq i \leq n} P^g_i\:,\\
P^g_0 & \de & \bigcup_{0 \leq j \leq m} D^g_j\:,\\
D^g_0 & \de & \tri{u_{0,0}}{u_{0,1}}{u_{0,2}}\union\vri{u_{0,2}}{r_{0,\mathrm{t}}}{r_{0,\mathrm{b}}}\:,\\
D^g_j & \de & \bigcup_{1 \leq p < j} \left\{\rule[-1ex]{0cm}{2ex}\vrii{w_{j,p}}{d_{p,\mathrm{t}}}{d_{p,\mathrm{b}}},\vrii{w_{j,p}}{d_{j,\mathrm{t}}}{d_{j,\mathrm{b}}}\right\}\union\\[-1ex]
         && \bigcup_{1 \leq k \leq n_j} \left\{\rule[-1ex]{0cm}{2ex}\vrii{f_{j,k}}{d_{j,\mathrm{t}}}{d_{j,\mathrm{b}}},\vrii{f_{j,k}}{g_{h,\mathrm{t}}}{g_{h,\mathrm{b}}}\:|\: l_{j,k}=x_h\right\}\\[-3ex]
         && \hspace{.65\textwidth}\mbox{for}\quad 1 \leq j \leq m\:,\\
P^g_i & \de & \tri{v_{i,0}}{v_{i,1}}{v_{i,2}}\union\left\{\rule[-1ex]{0cm}{2ex}\vrii{v_{i,2}}{g_{i,\mathrm{t}}}{g_{i,\mathrm{b}}},\right.\\[-1ex]
         && \ \:\left.\rule[-1ex]{0cm}{2ex}\vrii{v_{i,4}}{y_{i,\mathrm{t}}}{y_{i,\mathrm{b}}},\vrii{v_{i,4}}{g_{i,\mathrm{t}}}{g_{i,\mathrm{b}}}\right\}\quad\mbox{for}\quad 1 \leq i \leq n\:.
\end{eqnarray*}
\smallskip

\noindent Since printing and copying in color was not universally
available when this paper was written, and to avoid confusion
resulting from the large number of vertices and edges, the graph in
Figure~\ref{fig:gramsey} uses certain conventions to represent colors,
vertices and edges as indicated on its right-hand side. For instance,
we use $r_3$ as a shortcut for the edge $\{r_{3,\mathrm{t}}\setcomma
r_{3,\mathrm{b}}\}$, where ``t'' marks the vertex at the top of the
edge and ``b'' the one at the bottom.

It immediately follows from the construction that there is a simple
\logspace\ transducer that computes $(G,A,E^r,E^g)$ from input $F$.  We
still have to show that this construction ensures that there is a
winning strategy for player I of \gpos$(F)$ iff there is a
winning strategy for player Red of \gramsey$(G,A,E^r,E^g)$.

\begin{obs}\label{obs:construction}
By having a closer look at the construction of $(G,A,E^r,E^g)$, we
observe that edges $r_i$ for $i=0,\ldots ,n$ can only be chosen by
player Red. For instance, if player Green would color $r_2$ in green,
this would complete a green subgraph made of $v_{2,0}\setcomma
v_{2,1}\setcomma v_{2,2}\setcomma r_{2,\mathrm{t}}\setcomma$ and
$r_{2,\mathrm{b}}$ that would be isomorphic to $A$, which is forbidden
according to Definition~\ref{def:gramsey}. Similarly, all edges $d_j$
and $g_i$ for $j=1,\ldots ,m$ and $i=1,\ldots ,n$ can only be chosen
by player Green. Only the remaining edges $y_i$ for $i=1,\ldots ,n$
can initially be chosen by both players. However, once player Red
has chosen $y_i$ for some $i=1,\ldots ,n$, he cannot choose edge $r_i$
anymore, but Green can still play $g_i$, and vice versa for the
reversed roles of Red and Green. Thus, for each triple $r_i\setcomma
y_i\setcomma g_i$ for $i=1,\ldots ,n$, the first player to move has
the option to choose $y_i$, but the second player that colors an edge
in that triple can only select, in the case of Red, edge $r_i$, and in
the case of Green, edge $g_i$, once the middle edge $y_i$ has been
occupied. The remaining third edge of the triple always has to stay
uncolored. In other words, each player can color one edge of each
triple, but only the first to consider that particular triple has the
possibility to occupy the central edge, the player coming second being
left with the option to color his respective border edge once the
other player effectively has already chosen the central edge.
\end{obs}

\noindent Note that Green can only choose a single edge among edges $d_j$
because of the green connections through vertices
$w_{j,p}$. Additionally, Green cannot select both an edge $d_j$ and an
edge $g_i$ when the variable $x_i$ appears as literal $l_{j,k}$ in
conjunct $C_j$ of $F$ because of the connection through $f_{j,k}$.

\begin{fact}\label{fact:winning}
As a result of what has been said above, player Red can color
altogether exactly $n+1$ edges: namely $n$ edges out of the
$r_i\setcomma y_i$ for $i=1,\ldots ,n$, with no edges $r_i$ and $y_i$
from the same triple, plus $r_0$.  Player Green can either color
altogether at most $n$ or at most $n+1$ edges: namely at most $n$
edges out of the $y_i\setcomma g_i$ for $i=1,\ldots ,n$, with no edges
$y_i$ and $g_i$ from the same triple, plus at most one edge among
edges $d_j$, depending on the combination of edges $g_i$ and edge
$d_j$ colored in green and on which edges $g_i$ player Red could
previously ``force'' player Green to color in green by occupying the
corresponding edges $y_i$.
\end{fact}

\begin{lem}\label{lemma:npo}
For player Green being able to color less than $n+1$ edges means that
player Red has at least one more edge free to color at the end, so Red
can win, whereas for Green to be able to color exactly $n+1$ edges
means that both players color the same number of edges, and since Red
started, Green wins the game.
\end{lem}

\proof\ Follows immediately from Fact~\ref{fact:winning}. \halmos

\bigskip

\noindent We constructed $(G,A,E^r,E^g)$ in such a way as to constrain
the moves of the players in spite of their apparent freedom to choose
an uncolored edge, by punishing moves that are {\em illegitimate}.
Illegitimate moves are those that do not follow the complementary
notion of {\em legitimate\/} play, which is defined as follows:

\begin{defn}{\rm Legitimate play:}\label{def:legitimate} We call a 
\gramsey$(G,A,E^r,E^g)$ game sequence\/ {\em legitimate} iff it has
the following form: For moves $q=1,2,\ldots ,n$ both players choose so
far uncolored edges $y_{i_q}$, where $i_q\in\{1,\ldots,n\}$.  For
moves $q=n+1,n+2,\ldots ,2n+2$ player Red chooses colorable edges of
type $r_i$ with $i\in\{0,\ldots,n\}$, and player Green chooses
colorable edges of type $g_i$, with $i\in\{1,\ldots ,n\}$ and, if
possible, one edge $d_j$.
\end{defn}

\noindent We observe that the first $n$ moves of a legitimate game
sequence played on \gramsey$(G,A,E^r,E^g)$ mimic those of \gpos$(F)$
in an obvious way: Player Red on move $q$ chooses edge $y_{i_q}$ where
player I chooses variable $x_{i_q}$, and similar for players Green and
II. Let us note that in our construction, an illegitimate move may put
the player who selects such a move into a less favorable position for
the rest of the play.

Once at least one edge in every triple $r_i\setcomma y_i\setcomma g_i$
for $i=1,\ldots ,n$ is colored, all remaining playable edges are\/
{\em uncontested}, that is, only one of the two players can color each
particular colorable edge that is left, or, said in another way, the
two players cannot take away edges from each other anymore.

\begin{defn}{\rm The racing phase:} 
We call the part of a game sequence until only uncontested moves
remain the\/ {\em racing phase} of the \gramsey\ game, because during
that phase the two players race to occupy the `right' edges $y_i$ that
ultimately will lead to the victory of one of them.
\end{defn}

\noindent After this racing phase, each player becomes preoccupied
with his own set of edges that are left to play for him alone and
tries to play a solitaire in it as long as possible. The order in
which Red plays his remaining colorable edges is irrelevant, and Green
can maximize the number of edges he can color by coloring all
remaining colorable edges $g_{i_q}$ and, if available, $y_{i_q}$, the
order being again irrelevant, with the possible addition of one edge
$d_j$, depending on which edges $g_{i_q}$ Green colors during a game
sequence.

\begin{lem}\label{lemma:djlast}
If Green can color some $d_j$ at one of his moves and also is able to
color every $g_i$ such that Red selected $y_i$ during the racing
phase, then Green could as well have chosen to color $d_j$ as his last
move.
\end{lem}

\proof\ Independently of the notion of legitimate play, it does not
matter during which move Green colors this one edge $d_j$, i.e., it
needs not to be his last move, but Green still must be able to color
every edge $g_i$ where Red colored edge $y_i$ during the racing
phase. On the one hand, if Green can color some $d_j$ at one of
his moves and also is able to color every $g_i$ such that Red
selected $y_i$ during the racing phase, then Green could as well have
chosen to color $d_j$ as his last move. On the other hand, if Green
chooses to color some $d_j$ at one of his moves such that he cannot
color every $g_i$ where Red selected $y_i$ during the racing
phase, then Green could as well have chosen to color one of these
edges $g_i$ instead of the edge $d_j$. \halmos

\bigskip

\noindent As remarked in Lemma~\ref{lemma:npo}, win or loss depends
only on the difference in the number of edges the two players can
color, but not on the particular edges they color. In the following,
we thus can assume without loss of generality that, if permitted at
all by the coloring of edges $g_i$, Green always colors any edge
$d_j$\/ {\em after} coloring all possible edges of type $y_i$
and~$g_i$.

We will also need the following weaker variant of legitimate play:

\begin{defn}{\rm Winner-legitimate play:}\label{def:winnerlegitimate} We
call a game sequence\/ {\em winner-legitimate} if the player with the
winning strategy always has chosen legitimate moves.
\end{defn}

\noindent The following key lemma will allow us to decide in each case
who can win the \gramsey\ game:

\begin{lem}\label{lemma:winning}
Consider the game situation just before move $2n+2$ in a
winner-legitimate game played on \gramsey$(G,A,E^r,E^g)$. Player I has a
winning strategy on the corresponding \gpos$(F)$ game iff there exists
no edge $d_j$ that Green can choose after coloring all possible edges
of type $g_i$.
\end{lem}

\proof\ The statement follows straightforward from the following
facts:
\begin{enumerate}
\item Observation~\ref{obs:gposwbg} which says that either one of the
two players of \gpos$(F)$ has a winning strategy,
\item Definition~\ref{def:winnerlegitimate} of winner-legitimate play on
\gramsey$(G,A,E^r,E^g)$ that forces the player with the winning
strategy to first color edges $y_i$,
\item Lemma~\ref{lemma:djlast} which says we can always assume
that Green first colors all possible edges $y_i$ and $g_i$ before
coloring any edge $d_j$,
\item Lemma~\ref{lemma:npo} and the previous fact, which together
imply that Red and Green color in sum $2n+1$ edges before Green colors
any edge $d_j$, and
\item the one-to-one correspondence remarked in Fact~\ref{fact:iff}
between,
\begin{enumerate}
\item in the case of \gpos$(F)$, for player I to win iff he succeeds
in playing some variable $x_i$ in each conjunct $C_j$, and,
\item in the case of \gramsey$(G,A,E^r,E^g)$, for the first
player, Red, to win iff he succeeds in playing some edges $y_i$
(note that there is an edge $y_i$ for each variable $x_i$ in $F$) so
that his opponent, player Green, later on can only choose edges $g_i$
in these particular triples, making it impossible for him to color any
edge $d_j$ (remember that there is an edge $d_j$ for each conjunct
$C_j$ in $F$, and that edges $g_i$ and $d_j$ are connected through
$f_{j,k}$ iff variable $x_i$ appears in conjunct $C_j$). \halmos
\end{enumerate}
\end{enumerate}

\bigskip

\noindent In the following, we show that player I can win \gpos$(F)$
iff player Red can win \gramsey$(G,A,E^r,E^g)$, which concludes the
proof.

\bigskip \noindent $(\Rightarrow)$ Assume that player I has a winning
strategy for \gpos$(F)$. We first claim that Red has a strategy for
\gramsey$(G,A,E^r,E^g)$ that wins any game in which Green plays
legitimately. The strategy consists of playing legitimately and
applying player I's winning strategy for \gpos$(F)$ during the racing
phase, via the correspondence between variables $x_{i_q}$ and edges
$y_{i_q}$. After the racing phase is over, that is in case of
legitimate play, after $n$ moves, player Red and Green alternate in
coloring edges of the two disjoint sets of uncontested moves of each
player. The uncontested moves of Red consist in edge $r_0$ and in all
edges $r_{i_q}$ such that Green played $y_{i_q}$ during the racing
phase, so Red altogether colors $n+1$ edges during the game. The
uncontested moves of Green can consist in all edges $g_{i_q}$ such
that Red played $y_{i_q}$ during the racing phase, so Green altogether
colors $n$ edges before coloring any edge $d_j$. Because of
Lemma~\ref{lemma:djlast}, we can assume without loss of generality
that Green colors any edge $d_j$ only after coloring all possible
edges $g_i$ and $y_i$. In sum, this makes $2n+1$ moves for both
players before Green colors any edge $d_j$. However, it is easy to
check that after move $2n+1$, that is, when it would be again Green's
turn to play, the sufficient conditions for Green to be unable to
color any edge $d_j$ after coloring all possible edges of type $g_i$
stated in Lemma~\ref{lemma:winning} do hold, and so Red wins, as
described in Lemma~\ref{lemma:npo}.

It remains to show that Red can also win if Green does not play
legitimately. Suppose Green makes any illegitimate move at some point,
when all previous play was legitimate or at least winner-legitimate. We show
that, whatever this move is, Red has a response such that the game
continues with no disadvantage to Red but with a possible disadvantage
for Green. In the following, we examine all possible illegitimate moves
by Green. In light of Lemma~\ref{lemma:djlast}, we can always
assume without loss of generality that coloring any edge $d_j$ is the
last move player Green makes, which is in accordance to legitimate play,
so the only illegitimate move that Green is free to make is to color
some edge $g_{i_q}$ during the racing phase when a legitimate move for
him would be to color some edge $y_{i_q}$ instead. Red responds by
playing {\em as if Green just had chosen $y_{i_q}$}, and the game
continues in winner-legitimate way as if no illegitimate move had been
played. Red is none the worse off since the net result after the
racing phase is that Green voluntarily colored at least one edge
$g_{i_q}$ more than necessary, thus making it only harder for Green to
find an edge $d_j$ that can be colored in the last move. Red's play
is totally unaffected by Green's illegitimate play, and so again, after
move $2n+1$ the sufficient conditions for Green to be unable to color
any edge $d_j$ after coloring all possible edges of type $g_i$ stated
in Lemma~\ref{lemma:winning} do hold, and so Red wins, as described in
Lemma~\ref{lemma:npo}.

Thus, no matter what illegitimate moves Green makes, Red can win. This
completes the proof of the $(\Rightarrow)$ part.

\bigskip \noindent $(\Leftarrow)$ Assume that player II has a winning
strategy for \gpos$(F)$. We first claim that Green has a strategy for
\gramsey$(G,A,E^r,E^g)$ that wins any game in which Red plays
legitimately. The strategy again consists of playing legitimately and
applying player II's winning strategy for \gpos$(F)$ during the racing
phase, via the correspondence between variables $x_{i_q}$ and edges
$y_{i_q}$. After the racing phase is over, that is in case of legitimate
play, after $n$ moves, player Red and Green again alternate in
coloring edges of two disjoint sets of uncontested moves of each
player. The uncontested moves of Red consist in edge $r_0$ and in all
edges $r_{i_q}$ such that Green played $y_{i_q}$ during the racing
phase, so Red again altogether colors $n+1$ edges during the game. The
uncontested moves of Green consist in all edges $g_{i_q}$ such that
Red played $y_{i_q}$ during the racing phase, so Green altogether
colors $n$ edges before he attempts to color some edge $d_j$. After
move $2n+1$, that is, when it is again Green's turn to play, the
necessary conditions for Green to be unable to color any edge $d_j$
after coloring all possible edges of type $g_i$ stated in
Lemma~\ref{lemma:winning} do not hold, therefore Green is able to
color some $d_j$ and so Green wins, there being no edge left to color
for Red after move $2n+2$, as described in Lemma~\ref{lemma:npo}.

It remains to show that Green can also win if Red does not play
legitimately. Suppose Red makes any illegitimate move at some point, when
all previous play was legitimate or at least winner-legitimate. We show
that, whatever this move is, Green has a response such that the game
continues with no disadvantage to Green but also with no advantage for
Red.

In the following, we examine all possible illegitimate moves by Red,
assuming in each case that all previous play was legitimate or at least
winner-legitimate.

\bigskip {\em Case\/} 1: During the racing phase, Red plays some edge
$r_{i_q}$, where $i_q\in\{1,\ldots,n\}$, when a legitimate move for
him would be to color some edge $y_{i_q}$ instead. In this case Green
responds by playing {\em as if Red just had chosen $y_{i_q}$}, and the
game continues in winner-legitimate way, Green's strategy staying just
as if no illegitimate move had been played. Clearly, Green is none the
worse off by Red's choice since Red's illegitimate play will not
influence Green's capability to color some $d_j$ as his last move. In
summary, Green's play and strategy can remain totally unaffected by
Red's illegitimate play. After move $2n+1$, that is, when it is again
Green's turn to play, the necessary conditions for Green to be unable
to color any edge $d_j$ after coloring all possible edges of type
$g_i$ stated in Lemma~\ref{lemma:winning} do not hold, therefore Green
is able to color some $d_j$ and so Green wins, there being no edge
left to color for Red after move $2n+2$, as described in
Lemma~\ref{lemma:npo}.

\bigskip {\em Case\/} 2: During the racing phase, Red plays edge
$r_0$, when a legitimate move for him would be to color some edge
$y_{i_q}$ instead. In this case, again, Green responds by playing {\em
as if Red just had chosen $y_{i_q}$}, and the game continues in
winner-legitimate way as if no illegitimate move had been played. There is
a small technicality to be observed for Green, since the triple
containing $y_{i_q}$ is not really uncontested yet, but Green has to
consider it as such, whereas Red does not care about it. At some later
point of the play, Red will choose to color $y_{i_q}$ or $r_{i_q}$. To
see that this effectively will happen, remember that Red needs to
color either edge $r_i$ or $y_i$ in every triple in any game, as
remarked in Fact~\ref{fact:winning}. After Red's coloring of
$y_{i_q}$ or $r_{i_q}$, Green has to differentiate between the
following two subcases:

\bigskip {\em Subcase\/} 2a: If Red's move ends the racing phase,
Green continues as if the racing phase had ended already at Green's
previous move and as if Red had only now chosen to color $r_0$ and
previously had colored $y_{i_q}$ when Red actually had colored
$r_0$. Clearly, the game situations of the game really played so far
and the game in which Red's moves in question would have been played
the other way around are identical after Red's move, so the rest of
the play can continue as if no exchange of Red's two moves had ever
occurred. In summary, Green's play and strategy can remain totally
unaffected by Red's illegitimate play. After move $2n+1$, that is, when
it is again Green's turn to play, the necessary conditions for Green
to be unable to color any edge $d_j$ after coloring all possible edges
of type $g_i$ stated in Lemma~\ref{lemma:winning} do not hold,
therefore Green is able to color some $d_j$ and so Green wins, there
being no edge left to color for Red after move $2n+2$, as described in
Lemma~\ref{lemma:npo}.

\bigskip {\em Subcase\/} 2b: If Red's move does not end the racing
phase, there must be at least one triple that is not uncontested left
to color at Green's turn, with an edge $y_{i_{q'}}$. Green responds by
playing {\em as if Red just had chosen $y_{i_{q'}}$}, and the game
again continues in winner-legitimate way as if no illegitimate move had
been played. That is, Green only replaces the now uncontested
$y_{i_q}$ by the not yet uncontested $y_{i_{q'}}$, and we are again in
the situation of Case~2 from above, the only difference being that
Green now imagines that Red played $y_{i_{q'}}$ instead of $r_0$,
forgetting about any special treatment of $y_{i_q}$.

\bigskip
\noindent Thus, no matter what illegitimate moves Red
makes, Green can win. This completes the proof of the $(\Leftarrow)$
part and also the proof of Theorem~\ref{theo:gramsey}. \halmoseol

\sproof{theo:gmisereramsey}

Membership is identical to its treatment in the proof of
Theorem~\ref{theo:gramsey}, again following from
Lemma~\ref{lemma:membership}.  Hardness follows from
Definition~\ref{def:gmisereramsey} as well as
Theorem~\ref{theo:gramsey} and the construction in its proof, which
effectively makes sure that \gmisereramsey$(G,A,E^r,E^g)$ never ends
in a tie, by forcing $(G,E^r,E^g)\rightarrow A$. Indeed, as described
in Observation~\ref{obs:construction}, the construction features,
among others, $n$ triples $r_i\setcomma y_i\setcomma g_i$ for
$i=1,\ldots ,n$, such that coloring more than {\em one\/} edge in any
triple would end a \gmisereramsey\ game for that player. Since each
triple contains three edges but there are only two player, no
\gmisereramsey\ game will ever end in a tie because all edges have
been occupied. Therefore, Corollary~\ref{cor:misere} ensures that
\gmisereramsey$(G,A,E^r,E^g)$ will have the same winning strategy as
\gramsey$(G,A,E^r,E^g)$ for one of its players, and the proof of
Theorem~\ref{theo:gramsey} carries over. \halmoseol

\sproof{theo:gramseyplus}

The membership of \gramseyplus\ in \pspace\ follows again from
Lemma~\ref{lemma:membership}.

For the hardness part, a careful analysis of the proof of
Theorem~\ref{theo:gramsey} reveals that we can reuse the reduction of
that proof to show the \pspacec{}ness of \gramseyplus. Indeed, all
arguments go through even when both players are allowed to color more
than one edge per move. The difficulty here lies in the analysis of
the cases when the opponent plays illegitimately, as described below:

Let us assume that player I has a winning strategy and player Red has
so far played according to it, as explained in the proof of
Theorem~\ref{theo:gramsey}. In addition to everything which has
already been said there, we notice that if all previous play was
legitimate or at least winner-legitimate and player Green colors at
least two edges in his current move, the best he can hope for is that
he will have colored {\em one\/} edge $d_j$ at the end of the game. At any
rate, the number of edges Green has left to color altogether after his
move will decrease by at least two. Red is none the worse off by
Green's move and actually just needs to continue to choose {\em one\/}
uncolored edge $r_i$ after the other per move to win, since there is
no urge now to force Green to color edges $g_i$. Conversely, let us
assume player II has a winning strategy and player Green so far
followed it.  If player Red colors at least two edges in some move,
the best he can hope for is that this will disable Green to color an
edge $d_j$ as his last move, so Green has only one edge less left to
color during the rest of the game. However, the number of edges Red
has left to color decreases by the number of edges he colored, so
Green is none the worse off and still wins, now without having to
worry to color one additional edge $d_j$ at the end of the game. Thus,
Green just needs to choose one uncolored edge $g_i$ after the other
per move to win, since he is the second player and he has now at least
as many edges left to color as Red.  \halmoseol

\scor{cor:avoidrestriction}

Follows directly from the bow-tie construction of $A$ in the proof of
Theorem~\ref{theo:gramsey}. \halmoseol

\sproof{theo:gweakachieveramsey}
\begin{figure}[!t]
\begin{center}\epsfig{file=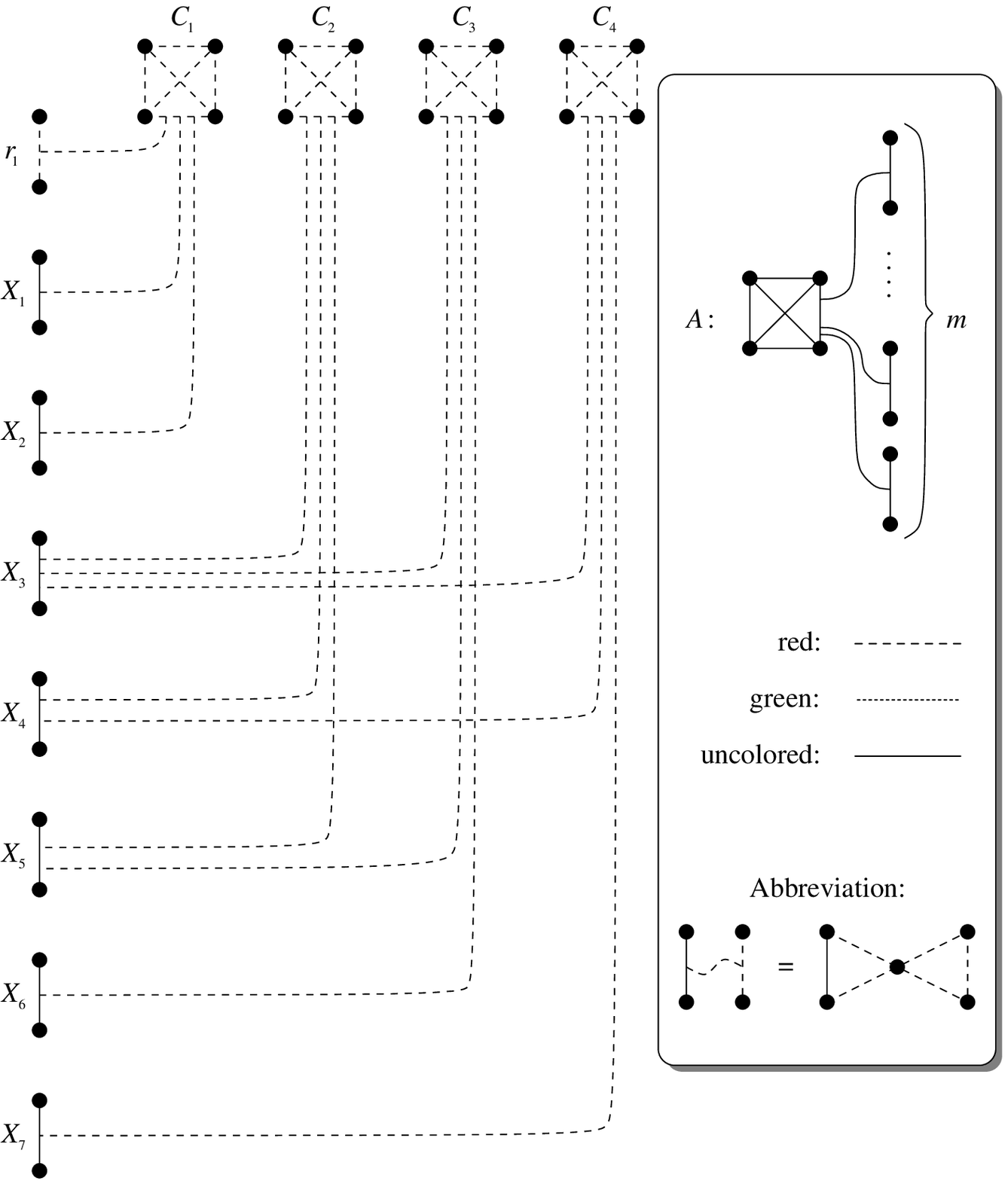,width=.94\textwidth}\end{center}
\boxcaption{Instance of the \gweakachieveramsey\ game
corresponding to the \gposdnf\ input formula $F=(x_1\And x_2)\Or
(x_3\And x_4\And x_5)\Or (x_3\And x_5\And x_6)\Or (x_3\And x_4\And
x_7)$, which features a winning strategy for player II. The number $m$
is the size of the largest disjunct in $F$.}{fig:gweakachieveramsey}
\end{figure}

We show that there is a \logspace\ reduction from \gposdnf\ to the
game \gweakachieveramsey.  
By a result of T.~\citet{Schaefer78} showing that \gposdnf\ is
\pspacec\ and the obvious membership of the \gweakachieveramsey\ game in
\pspace\ (again following from Lemma~\ref{lemma:membership}), the
result follows.

The definition of \gposdnf\ is restated here:
\begin{defn}[T.~\citet{Schaefer78}] \gposdnf$(F)$ We are given a positive DNF formula
$F$. A move consists of choosing some variable of $F$ which has not
yet been chosen. Player I starts the game. The game ends after all
variables of $F$ have been chosen. Player I wins iff $F$ is\/ {\tt true}
when all variables chosen by player I are set to\/ {\tt true} and all
variables chosen by player II are set to\/ {\tt false}.  In other words,
player I wins iff he succeeds in playing all variables in at least one
disjunct.
\end{defn}

\begin{theo}[T.~\citet{Schaefer78}] ~\\
\hspace*{\parindent}\gposdnf\ is \pspacec.
\end{theo}

\noindent We next describe the \logspace\ reduction from \gposdnf$(F)$
to \gweakachieveramsey$(G,A,E^r,E^g)$.
Figure~\ref{fig:gweakachieveramsey} shows a small example. The exact
definition follows: Let a positive DNF formula $F$ be given. Assume
without loss of generality that $F = D_1 \Or \ldots \Or D_q$ where
each disjunct $D_j$ is a conjunction of $n_j$ positive literals, that
is, $D_j = l_{j,1} \And \ldots \And l_{j,n_j}$ where $l_{j,k} \in
\{x_1, \ldots, x_n\}$ and all $n$ variables appear at least once in
$F$.  We then define the graphs $G\de(V,E)$, $A\de(V^A,E^A)$ and the
edge-sets $E^r$, $E^g$, by \rem{no splitting of formulas over
page-breaks!}
\begin{eqnarray*}
V   & \de & \bigcup_{0 \leq i \leq n} X_i\:,\\
X_0 & \de & \bigcup_{0 \leq j \leq q} C_j\:,\\
C_0 & \de & \bigcup_{1 \leq k \leq p} \{r_{k,\mathrm{t}}\setcomma r_{k,\mathrm{b}}\}\:,\\
p   & \de & m\:-\:\min_{1 \leq j \leq q} \{n_j\}\:,\\
m   & \de & \max_{1 \leq j \leq q} \{n_j\}\:,\\
C_j & \de & \{u_{j,0}\setcomma u_{j,1}\setcomma u_{j,2}\setcomma u_{j,3}\}\union\: \bigcup_{1 \leq i < m} \{v_{i,j}\}\quad\mbox{for}\quad 1 \leq j \leq q\:,\\
X_i & \de & \{x_{i,\mathrm{t}}\setcomma x_{i,\mathrm{b}}\}\quad\mbox{for}\quad 1 \leq i \leq n\:,\\
E   & \de & \bigcup_{0 \leq i \leq n} P_i\:,\\
P_0 & \de & \bigcup_{0 \leq j \leq q} D_j\:,\\
D_0 & \de & \bigcup_{1 \leq k \leq p} \left\{\rule[-1ex]{0cm}{2ex}\{r_{k,\mathrm{t}}\setcomma r_{k,\mathrm{b}}\}\right\}\:,\\
D_j & \de & \left\{\rule[-1ex]{0cm}{2ex}\{u_{j,0}\setcomma u_{j,1}\}\setcomma \{u_{j,0}\setcomma u_{j,2}\}\setcomma \{u_{j,0}\setcomma u_{j,3}\}\setcomma\right.\\[-1ex]
       && \left.\rule[-1ex]{0cm}{2ex}~~\{u_{j,1}\setcomma u_{j,2}\}\setcomma \{u_{j,1}\setcomma u_{j,3}\}\setcomma \{u_{j,2}\setcomma u_{j,3}\}\right\}\union\\[-1ex]
       && \bigcup_{1 \leq k \leq n_j} \left\{\rule[-1ex]{0cm}{2ex}\{u_{j,0}\setcomma v_{i,j}\}\setcomma \{u_{j,1}\setcomma v_{i,j}\}\setcomma\right.\\[-3ex]
       && \left.\rule[-1ex]{0cm}{2ex}\hspace*{3.75em}\{v_{i,j}\setcomma x_{i,\mathrm{t}}\}\setcomma \{v_{i,j}\setcomma x_{i,\mathrm{b}}\}\:|\: l_{j,k}=x_i\right\}\union\\[-1ex]
       && \bigcup_{1 \leq k \leq m-n_j} \left\{\rule[-1ex]{0cm}{2ex}\{u_{j,0}\setcomma v_{n_j+k,j}\}\setcomma \{u_{j,1}\setcomma v_{n_j+k,j}\}\setcomma\right.\\[-3ex]
       && \left.\rule[-1ex]{0cm}{2ex}\hspace*{5em}\{v_{n_j+k,j}\setcomma r_{k,\mathrm{t}}\}\setcomma \{v_{n_j+k,j}\setcomma r_{k,\mathrm{b}}\}\right\}\qquad\mbox{for}\quad 1 \leq j \leq q\:,\\
P_i & \de & \left\{\rule[-1ex]{0cm}{2ex}\{x_{i,\mathrm{t}}\setcomma x_{i,\mathrm{b}}\}\right\}\qquad\mbox{for}\quad 1 \leq i \leq n\:,\\
V^A & \de & \{a_0\setcomma a_1\setcomma a_2\setcomma a_3\}\union\: \bigcup_{1 \leq i < m} \{b_{i,0}\setcomma b_{i,1}\setcomma b_{i,2}\}\:,\\
E^A & \de & \left\{\rule[-1ex]{0cm}{2ex}\{a_{0}\setcomma a_{1}\}\setcomma \{a_{0}\setcomma a_{2}\}\setcomma \{a_{0}\setcomma a_{3}\}\setcomma \{a_{1}\setcomma a_{2}\}\setcomma \{a_{1}\setcomma a_{3}\}\setcomma \{a_{2}\setcomma a_{3}\}\right\}\union\\[-1ex]
       && \bigcup_{1 \leq i \leq m} \left\{\rule[-1ex]{0cm}{2ex}\{a_{0}\setcomma b_{i,0}\}\setcomma \{a_{1}\setcomma b_{i,0}\}\setcomma \{b_{i,0}\setcomma b_{i,1}\}\setcomma \{b_{i,0}\setcomma b_{i,2}\}\setcomma \{b_{i,1}\setcomma b_{i,2}\}\right\}\:,\\
E^r & \de & P_0\:,\\
E^g & \de & \{\}\:.
\end{eqnarray*}
\smallskip

\noindent It immediately follows from the construction that there is a
simple \logspace\ transducer that computes $(G,A,E^r,E^g)$ from input
$F$.

Since the number of variables that can be chosen in \gposdnf\ is equal
to the number of edges that can be colored in \gweakachieveramsey, it
is easy to see that there is a one-to-one correspondence between
variables and edges. A winning strategy from \gposdnf\ is directly
translated into a winning strategy for the \gweakachieveramsey\ game
by coloring edge $X_i$ whenever variable $x_i$ of $F$ needs to be
chosen, and vice versa.

The graph $A$ that player Red has to complete in his color looks like
an `$m$-legged octopus', we therefore call it an `$m$-topus' in the
following. Each disjunct $D_j$ of the positive DNF formula $F$ is
mirrored by an $m$-topus already partly precolored in red such that
only the `feet-edges' of the $m$-topus that correspond to the
variables occurring in the disjunct are still uncolored.

In the following, we show that above construction ensures that there
is a winning strategy for player I of \gposdnf$(F)$ iff there is a
winning strategy for player Red of \gweakachieveramsey$(G,A,E^r,E^g)$.
The strategy consists in copying the winning strategy for
\gposdnf$(F)$ via the correspondence between variables $x_i$ and edges
$X_i$. It is easy to see that player I can play all variables in at
least one disjunct iff player Red can color the feet-edges of at
least one $m$-topus. Since player I wins the game \gposdnf$(F)$ iff he
can play all variables in at least one disjunct, and since player Red
wins the game \gweakachieveramsey$(G,A,E^r,E^g)$ iff he completes to
color the feet-edges of at least one $m$-topus, thereby building a
red subgraph isomorphic to $A$, this concludes the proof. \halmoseol

\sproof{theo:gachieveramseynt}

Since the construction used in the proof of
Theorem~\ref{theo:gweakachieveramsey} leaves no possibility open for
Green to construct a green subgraph isomorphic to $A$, we can
reinterpret the whole proof according to the rules of
\gachieveramseynt. It is easy to see that \gweakachieveramsey\ and
\gachieveramseynt\ are in fact the same game when the second player
cannot build a green subgraph isomorphic to $A$, and so the
\pspacec{}ness proof remains true if we replace every occurrence of
\gweakachieveramsey\ by an occurrence of \gachieveramseynt. Therefore,
the \pspacec{}ness of \gweakachieveramsey\ directly carries over to
the \gachieveramseynt\ game. \halmoseol

\scor{cor:achieverestriction}

Follows directly from the $m$-topus construction of $A$ in the proof
of Theorem~\ref{theo:gweakachieveramsey} and the restriction to DNF
formulas having at most 11 variables in each disjunct in the
\pspacec{}ness proof of \gposdnf\ (T.~\citet[Theorem 3.6 and Corollary
3.7]{Schaefer78}), which limits the maximum degree of the vertices in
the achievement graph $A$ to 14. \halmoseol

\sproof{theo:gachieveramsey}

Membership in \pspace\ again easily follows again from
Lemma~\ref{lemma:membership}. To show hardness, we will adapt the
reduction from the proof of Theorem~\ref{theo:gweakachieveramsey} for
the present proof.  Indeed, all we have to make sure is that player
Red has a winning strategy in \gweakachieveramsey\ iff Red also wins
the corresponding \gachieveramsey\ game. The modifications we describe
below ensure that, on the one hand, if player Red has a winning
strategy in \gweakachieveramsey\ and thus can construct a red subgraph
isomorphic to graph $A$, this winning strategy will carry over to
\gachieveramsey\ without change.  On the other hand, if player Green
can prevent Red from constructing such a red subgraph, player Green
has a winning strategy in \gweakachieveramsey\ but not (yet) in
\gachieveramsey, so for the (modified) latter game we need to add a
gadget that makes sure Green can build a green subgraph isomorphic to
$A$, of course only in case Red cannot build one earlier in red. This
would make sure that Green would be given a winning strategy in
\gachieveramsey\ in case Green had a winning strategy in
\gweakachieveramsey.

\begin{figure}
\begin{center}\epsfig{file=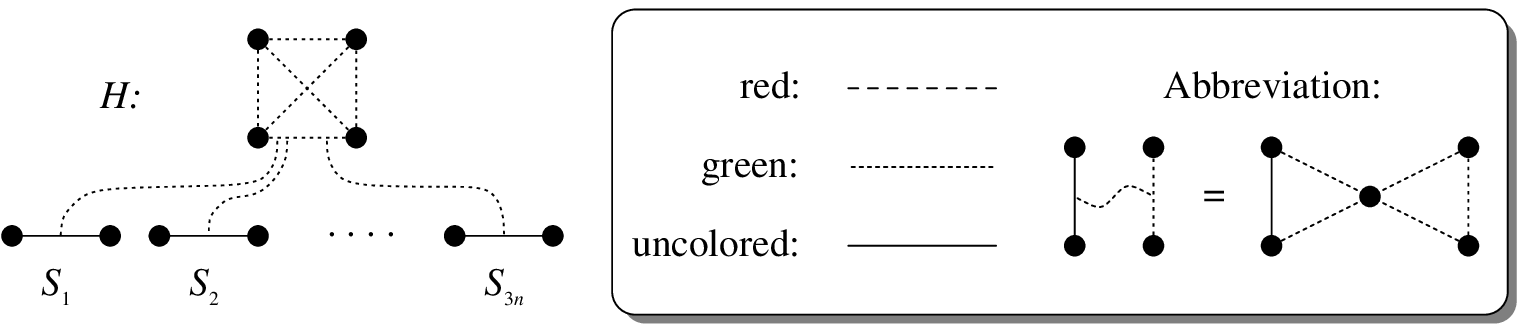,width=\textwidth}\end{center}
\boxcaption{$3n$-topus gadget that must be added to the construction
in Figure~\ref{fig:gweakachieveramsey} besides the redefinition of $m$
to the number of variables in order to transform the reduction used in
the proof of Theorem~\ref{theo:gweakachieveramsey} to one that allows
to prove Theorem~\ref{theo:gachieveramsey}. Conventions are similar to
those in Figure~\ref{fig:gramsey}.}{fig:gachieveramsey}
\end{figure}

Let us first define the modified reduction in detail. Almost
everything is defined as in the proof of
Theorem~\ref{theo:gweakachieveramsey} besides the redefinition of $m$,
of $G$, and of $E^g$. The changes add the new graph $H\de(V^H,E^H)$,
namely a $3n$-topus partly precolored in green with all $3n$ feet-edges
still uncolored, as depicted in Figure~\ref{fig:gachieveramsey}, to
the initial game situation, increase the number of `legs' in $A$ from
the size of the largest clause to the number of variables $n$, and add
legs already precolored completely in red to each gadget $C_i$
corresponding to disjunct $D_i$ such that in case Red had a winning
strategy in \gweakachieveramsey, it can be reused without change in
the new \gachieveramsey\ game. Below we only state the formulas that
were changed with respect to the formulas in the proof of
Theorem~\ref{theo:gweakachieveramsey}:
\begin{eqnarray*}
m   & \rde & n\:,\\
G   & \rde & (V\cup V^H\setcomma E\cup E^H)\:,\\
V^H & \de &  \{h_0\setcomma h_1\setcomma h_2\setcomma h_3\}\union\: \bigcup_{1 \leq i < 3n} \{s_{i,0}\setcomma s_{i,1}\setcomma s_{i,2}\}\:,\\
E^H & \de & \left\{\rule[-1ex]{0cm}{2ex}\{h_{0}\setcomma h_{1}\}\setcomma \{h_{0}\setcomma h_{2}\}\setcomma \{h_{0}\setcomma h_{3}\}\setcomma \{h_{1}\setcomma h_{2}\}\setcomma \{h_{1}\setcomma h_{3}\}\setcomma \{h_{2}\setcomma h_{3}\}\right\}\union\\[-1ex]
       && \bigcup_{1 \leq i \leq 3n} \left\{\rule[-1ex]{0cm}{2ex}\{h_{0}\setcomma s_{i,0}\}\setcomma \{h_{1}\setcomma s_{i,0}\}\setcomma \{s_{i,0}\setcomma s_{i,1}\}\setcomma \{s_{i,0}\setcomma s_{i,2}\}\setcomma \{s_{i,1}\setcomma s_{i,2}\}\right\}\:,\\
E^g & \rde & \left\{\rule[-1ex]{0cm}{2ex}\{h_{0}\setcomma h_{1}\}\setcomma \{h_{0}\setcomma h_{2}\}\setcomma \{h_{0}\setcomma h_{3}\}\setcomma \{h_{1}\setcomma h_{2}\}\setcomma \{h_{1}\setcomma h_{3}\}\setcomma \{h_{2}\setcomma h_{3}\}\right\}\union\\[-1ex]
       && \bigcup_{1 \leq i \leq 3n} \left\{\rule[-1ex]{0cm}{2ex}\{h_{0}\setcomma s_{i,0}\}\setcomma \{h_{1}\setcomma s_{i,0}\}\setcomma \{s_{i,0}\setcomma s_{i,1}\}\setcomma \{s_{i,0}\setcomma s_{i,2}\}\right\}\:.
\end{eqnarray*}
\smallskip

\noindent Similar to the proof of Theorem~\ref{theo:gramsey}, we will
use $S_i$ as a shortcut for the edge $\{s_{i,1}\setcomma
s_{i,2}\}$. The new $(G,A,E^r,E^g)$ is constructed as in the proof of
Theorem~\ref{theo:gramsey} in such a way as to constrain the moves of
the players in spite of their apparent freedom to choose an uncolored
edge, by punishing moves that are {\em illegitimate}. Again, an
illegitimate move may put the player who selects such a move into a
less favorable position for the rest of the play and is defined via
the complementary notion of {\em legitimate\/} play:

\begin{defn}{\rm Legitimate play:}\label{def:alegitimate} We call a 
\gachieveramsey$(G,A,E^r,E^g)$ game sequence\/ {\em legitimate} iff it
has the following form: For moves $q=1,2,\ldots ,n$ both players
choose so far uncolored edges $X_{i_q}$, where $i_q\in\{1,\ldots,n\}$.
In case Red builds a red subgraph isomorphic to $A$ during the first
$n$ moves, the game ends, in accordance with the rules of the
\gachieveramsey\ game. In case Red does not build a red subgraph
isomorphic to $A$ during the first $n$ moves, the game continues for
at most $2n$ more moves during which both players take turns to color edges
$S_i$ in the added $3n$-topus $H$.
\end{defn}

\noindent The following weaker variant of legitimate play corresponds
to its definition in the proof of Theorem~\ref{theo:gramsey}:

\begin{defn}{\rm Winner-legitimate play:} We
call a game sequence\/ {\em winner-legitimate} if the player with the
winning strategy always has chosen legitimate moves.
\end{defn}

\noindent As in the proof of Theorem~\ref{theo:gramsey}, the following
notion will be of help:

\begin{defn}{\rm The racing phase:}
We call the first $n$ moves of a game sequence the\/ {\em racing
phase} of the \gachieveramsey\ game, because during that phase the two
players race to occupy the `right' edges $X_i$ that ultimately will
lead to the victory of one of them.
\end{defn}

\noindent Let us observe that the moves of the racing phase of a
legitimate game sequence played on \gachieveramsey$(G,A,E^r,E^g)$
mimic those of \gposdnf$(F)$ in an obvious way: Player Red on move $q$
chooses edge $X_{i_q}$ where player I chooses variable $x_{i_q}$, and
similar for player Green and player~II.

We will show that this modified reduction will make sure that there is
a winning strategy for player I of \gposdnf$(F)$ iff there is a
winning strategy for player Red of \gachieveramsey$(G,A,E^r,E^g)$:

\bigskip \noindent $(\Rightarrow)$ Assume that player I has a winning
strategy for \gposdnf$(F)$. We first claim that Red has a strategy for
\gachieveramsey$(G,A,E^r,E^g)$ that wins any game in which Green plays
legitimately. The strategy consists of playing legitimately and
applying player I's winning strategy for \gposdnf$(F)$, via the
correspondence between variables $x_{i_q}$ and edges $X_{i_q}$. Player
Red will be able to complete a red subgraph isomorphic to $A$ and thus
win, the argument being the same as in the proof of
Theorem~\ref{theo:gweakachieveramsey}.

It remains to show that Red can also win if Green does not play
legitimately. Suppose Green makes any illegitimate move at some point,
when all previous play was legitimate or at least
winner-legitimate. We show that, whatever this move is, Red has a
response such that the game continues with no disadvantage to Red but
also with no advantage for Green. The only illegitimate move that
Green is free to make is to color some edge $S_{i_q}$ during the
racing phase when a legitimate move for him would be to color some
edge $X_{i_q}$ instead. Red responds by playing {\em as if Green just
had chosen $X_{i_q}$}, and the game continues in winner-legitimate way
as if no illegitimate move had been played. If Green later during the
game, say at move $q'$, chooses to actually color this edge $X_{i_q}$,
Red continues as if Green had colored $X_{i_q}$ already at move $q$
and another uncolored edge $X_{i_{q'}}$ at the present move. Red is
none the worse off since the net result after the racing phase is that
Green voluntarily renounced to color some edges $X_{i}$, thus making
it even easier for Red to complete a red subgraph isomorphic to
$A$. Since Red needed at most $\lceil n/2\rceil$ moves to construct a
red subgraph isomorphic to $A$, Green cannot yet have finished to
color the necessary $n$ feet-edges of $H$ that are needed to construct
a green subgraph isomorphic to $A$. Altogether, Red's play is totally
unaffected by Green's illegitimate play, and so again, after at most
$n$ steps, Red will be able to complete a red subgraph isomorphic to
$A$ and thus Red again will be able to win the game.

Thus, no matter what illegitimate moves Green makes, Red can win. This
completes the proof of the $(\Rightarrow)$ part.

\bigskip \noindent $(\Leftarrow)$ Assume that player II has a winning
strategy for \gposdnf$(F)$. We first claim that Green has a strategy
for \gachieveramsey$(G,A,E^r,E^g)$ that wins any game in which Red
plays legitimately. The strategy again consists of playing
legitimately and applying player II's winning strategy for
\gposdnf$(F)$ during the racing phase, via the correspondence between
variables $x_{i_q}$ and edges $X_{i_q}$. After the racing phase is
over, that is in case of legitimate play, after $n$ moves, Green will
have hindered Red from constructing a red subgraph isomorphic to $A$,
the argument being the same as in the proof of
Theorem~\ref{theo:gweakachieveramsey}. It is easy to see that from
then on, Red cannot hinder Green from building a green subgraph
isomorphic to $A$ by coloring $n$ feet-edges of the $3n$-topus $H$, and so
Green will win the game at move $3n$.

It remains to show that Green can also win if Red does not play
legitimately. Suppose Red makes any illegitimate move at some point,
when all previous play was legitimate or at least
winner-legitimate. We show that, whatever this move is, Green has a
response such that the game continues with no disadvantage to Green
but also with no advantage for Red. The only illegitimate move that
Red is free to make is to color some edge $S_{i_q}$ during the racing
phase when a legitimate move for him would be to color some edge
$X_{i_q}$ instead. Green responds by playing {\em as if Red just had
chosen $X_{i_q}$}, and the game continues in winner-legitimate way as
if no illegitimate move had been played. If Red later during the game,
say at move $q'$, chooses to actually color this edge $X_{i_q}$, Green
continues as if Red had colored $X_{i_q}$ already at move $q$ and
another uncolored edge $X_{i_{q'}}$ at the present move. Green is none
the worse off since the net result after the racing phase is that Red
voluntarily renounced to color some edges $X_{i}$, thus making it even
easier for Green to hinder Red from completing a red subgraph
isomorphic to $A$. Since only $n$ moves have been played so far, Red
will have colored at most $\lceil n/2\rceil$ edges $S_i$ in the
$3n$-topus $H$, and so there are still more than $3n-\lceil
n/2\rceil\geq 2n$ uncolored feet-edges of $H$ that Green can color to
complete a green subgraph isomorphic to $A$. Again, it is easy to see
that Green will win the game at move $3n$.

\bigskip
\noindent Thus, no matter what illegitimate moves Red makes, Green can
win. This completes the proof of the $(\Leftarrow)$ part and also the
proof of Theorem~\ref{theo:gachieveramsey}. \halmoseol

\section{Further complexity results for graph Ramsey games}\label{sec:furtherres}

We next study the complexities of two natural
generalizations of \gramsey. 

The first game, \gmultiramsey{n}, is based on generalizing
Definitions~\ref{def:classic-ramsey-nr} and \ref{def:arrowing} to more
arguments. Indeed, classic Ramsey numbers for more than two colors
such as \ramsey$(3,3,3)=17$ (\citet{Greenwood55}) also lead to
interesting combinatorial multi-player games. Note that
\ramsey$(3,3,3)=17$ is the only nontrivial case for which, to the best
of our knowledge, the answer is known so far for $n$-ary Ramsey
numbers with $n>2$ \cite{Radziszowski94}.

\begin{defn}{\gmultiramsey{n}$(G,A,E^1,\ldots,E^n)$:}\label{def:gmultiramsey}
We are given a graph $G=(V,E)$, another graph $A$, and $n\geq 2$
non-intersecting sets $\cup_{i=1}^n E^i \subseteq E$ that contain
edges initially colored in colors $c_1$ to $c_n$, respectively. One
after the other, the $n$ players select at each move one so-far
uncolored edge from $E$ and color it in their respective
color. However, all players are forbidden to choose an edge such that
$A$ becomes isomorphic to a monochromatic subgraph of $G$. Player one
starts. The first player unable to move, say player $l$, loses and
withdraws from the rest of the game. The winner of
\gmultiramsey{n}$(G,A,E^1,\ldots,E^n)$ is recursively defined as
follows:
\begin{description}
\item[Case $n=2$:] the winner is the sole player that is still in the
game.
\item[Case $n>2$:] \ the winner is the winner of the game
\gmultiramsey{n-1}$(G,A,\underbrace{\{\},\ldots,\{\}}_{n-1})$, where
the new player $i$ is the old player $(l+i-1\bmod n)+1$.
\end{description}
\end{defn}

\begin{corol}\label{cor:gmultiramsey} \
\gmultiramsey{n} is \pspacec.
\end{corol}

\proof\ Easily follows from Theorem~\ref{theo:gramsey} and the
observation that the recursion in Definition~\ref{def:gmultiramsey}
ends after $n-1$ steps. Membership in \pspace\ follows again from
Lemma~\ref{lemma:membership}.  \halmoseol

\mbox{}\\\noindent The second game, \gasymmetricramsey, is based on
the unchanged Definitions~\ref{def:classic-ramsey-nr} and
\ref{def:arrowing} but considers different avoidance graphs for the
two players. Again, only few results are known, for instance
\ramsey$(4,5)=25$ proved recently by \citet{McKay97} using a massive
amount of computing power, reportedly a cluster of workstations
running for more than 10 cpu-years (consult \cite{Radziszowski94} for
a survey on known small Ramsey numbers).

\begin{defn}{\gasymmetricramsey$(G,A^r,A^g,E^r,E^g)$:}
We are given a graph $G=(V,E)$, two more graphs $A^r$ and $A^g$, and
two non-intersecting sets $E^r \cup E^g \subseteq E$ that contain
edges initially colored in red and green, respectively. The two
players, Red and Green, select at each move one so-far uncolored edge
from $E$ and color it in red for player Red respectively in green for
player Green. However, player Red is forbidden to choose an edge such
that $A^r$ becomes isomorphic to a red subgraph of $G$, and player
Green is forbidden to choose an edge such that $A^g$ becomes
isomorphic to a green subgraph of $G$. It is Red's turn. The first
player unable to move loses.
\end{defn}

\begin{corol}\label{cor:gasymmetricramsey} \
\gasymmetricramsey\ is \pspacec.
\end{corol}

\proof\ Follows directly from Theorem~\ref{theo:gramsey}. \halmoseol

\mbox{}\\\noindent Of course, appropriate combinations of
\gmultiramsey{n} and \gasymmetricramsey\ as well as similar variants
of avoidance$^+$ and achievement games are conceivable and are also
\pspacec, following from Corollaries~\ref{cor:gmultiramsey}
and \ref{cor:gasymmetricramsey}.

\section{Solving concrete instances}\label{sec:concrete}

In Section~\ref{sec:proofs}, we proved that deciding graph Ramsey games
is \pspacec. Besides determining the theoretical computational
resources needed to solve a game's decision problem asymptotically, it
is possible to differentiate the following four levels of how to solve
a particular instance of a combinatorial game in practice:

\begin{defn}\label{def:solved}{\rm Levels of solving a 
combinatorial game in practice \cite{Allis94,Schaeffer96a}:}
\begin{description}
\item[Ultra-weakly solved:] The game-theoretic value for the initial
position has been determined.
\item[Weakly solved:] The game is ultra-weakly solved and a strategy
exists for achieving the game-theoretic value from the opening
position, assuming reasonable computing resources.
\item[Strongly solved:] For all possible positions, a strategy is
known for determining the game-theoretic value for both players,
assuming reasonable computing resources.
\item[Ultra-strongly solved:] For all positions in a strongly solved
game, a strategy is known that improves the chances of achieving more
than the game-theoretic value against a fallible opponent.
\end{description}
\end{defn}

\noindent Let us look at what can be achieved in the case of concrete
instances of \gramsey\ and \gramseyplus, starting with Sim, that is,
$\gramsey(K_{\scriptsize \ramsey(3,3)},K_3,\{\},\{\})$.

\subsection{Strongly solving Sim}

In accordance with Definition~\ref{def:gramsey}, we assume that player
Red always makes the first move, even if this is not necessarily so in
real games, in particular in our implementation of Sim, where
player Red is always the human player, independently of who starts. As
mentioned in Section~\ref{sec:defs}, in Sim the longest game sequence
lasts 15 moves. If none of the two players builds a monochromatic
triangle in the first 14 steps, it is the first player, Red, whose
turn it would be to make the 15$^{th}$, final move. That move,
however, for certain is fatal, for there must be a monochromatic
triangle according to the party-puzzle result from Ramsey theory
mentioned in Section~\ref{sec:defs} (see
Figure~\ref{fig:r33e6}). Intuitively, if both players can delay
building a triangle up to the 15$^{th}$ move, Red will loose, so Green
might have a winning strategy in Sim.

Strongly solving a game requires that the complete game tree is
known. For most games this tree is far too large to be generated and
evaluated backwards from the terminal game positions to determine this
strategy. Is it possible to generate the whole game tree for Sim?

In order to approximate the size of Sim's game tree, let us assume for
the moment that all play sequences will last 15 steps, thus including
successors of terminal positions that actually are impossible in real
games. Proceeding with this assumption, we can construct a game tree
with 15 alternatives on the first level, $15\times14$ on the second,
$15\times14\times13$ on the third and so on. That makes altogether
$15!$, around $1.3\times10^{12}$, leaves in the last level. We would
need memory in the order of 150 Gigabytes for a search tree of that
size even if we could manage to use only one bit for each
position. While that seems feasible, we wondered whether we could do
with less.

\begin{obs}\label{obs:isomorphism}
By looking at an arbitrary game sequence, we immediately
notice that at his starting move the first player actually has,
instead of 15 possibilities, only one possibility to choose from,
because all game positions are isomorphic at this level modulo a
permutation of the six vertices. Similarly, the second player does not
really choose between 14 edges, but instead can either select an edge
with a vertex in common to the edge colored by the first player, or
unconnected to that edge, summing up to two possibilities.
\end{obs}

\noindent It is easy to see that this isomorphism between game
positions dramatically reduces the size of the directed acyclic graph
corresponding to a compressed version of the fully expanded game
tree. We found that the number of non-isomorphic legal game positions,
including positions containing monochromatic triangles, was thus
reduced to a mere 3728. Actually, it is not even necessary to save the
positions that contain monochromatic triangles, since they can be
easily detected during play. Thus, the number can be further reduced
to 2309 positions. We defined a normalization based on choosing the
smallest member from the set of isomorphic positions in a convenient
lexicographic ordering as the representative of the isomorphism class
to convert the game positions in the numerical range [0, $3^{15}-1$]
to these 2309 positions. In our implementation, we stored these
positions in a hash table with $2^{12}=4096$ entries that can be
downloaded as a whole by the Java applet mentioned in
footnote~\ref{foot:java} so that no further contact with the
server-side nor any lengthy computations are necessary during play.

A position in the game tree is labeled R-WIN if the first player, Red,
has a winning strategy from that position, and R-LOSS
otherwise. Terminal positions of the implemented mis\`ere variant of
Sim are labeled R-WIN if Green closed a triangle, and R-LOSS
otherwise. The positions that do not contain triangles are recursively
labeled from bottom-up as follows. If it is Red's turn, and at least
one of the positions following the current one in the game tree is
labeled R-WIN, then the current position is also labeled R-WIN, since
selecting that move would lead to Red's victory. If, however, all of
them are R-LOSS, then Green will be able to beat Red whatever move Red
chooses, so the current position must be labeled R-LOSS. If it is
Green's turn, at least one move leading to an R-LOSS position is enough
to label the current one R-LOSS, and only if all possibilities are
R-WIN must it be labeled R-WIN as well. Of course isomorphic game
positions have to be labeled only once, so it is enough to label the
2309 normalized positions in the directed acyclic graph. Our results
coincide with those of \citet{Mead74}:

\begin{theo}\label{theo:simloss}
The second player, Green, has a winning strategy in Sim =
\gramsey$(K_{\scriptsize \ramsey(3,3)},K_3,\{\},\{\})$.
\end{theo}

\noindent The proof of this statement being more or less a very long
enumeration of cases for the game \gmisereramsey$(K_{\scriptsize
\ramsey(3,3)},K_3,\{\},\{\})$, we do not include it here. It can
easily be reconstructed from the data available together with the
Java-applet on the author's home-page. Because of
Corollary~\ref{cor:misere}, \gmisereramsey$(K_{\scriptsize
\ramsey(3,3)},K_3,\{\},\{\})$ is equivalent to
\gramsey$(K_{\scriptsize \ramsey(3,3)},K_3,\{\},\{\})$, so the
statement is true.

Accordingly, the second player has a winning strategy which we
implemented in our Java applet.

Theorem~\ref{theo:simloss} established that we {\em ultra-weakly
solved\/} Sim in the sense of Definition~\ref{def:solved}.  However,
since we are also able to quickly generate the game-theoretic value of
all possible game positions from our data, we effectively both {\em
weakly\/} and {\em strongly solved\/} the Sim game.

An interesting problem that remained was to find a strategy when our
program plays as the first player, i.e., to additionally {\em
ultra-strongly solve\/} the game. Our solution is presented in the next
section.

\subsection{Ultra-strongly solving Sim}

We have already compressed the game tree and saved it in a hash
table. If the second player is perfect, the first player has no chance
to win. However, it may be extremely difficult for a human second
player to choose a perfect move at every single step, unless one
memorizes the directed acyclic graph with the 3728 non-isomorphic game
positions and additionally is able to identify the current position
with the corresponding normalized one. As soon as the human player by
mistake chooses a non-perfect move, the program is able to follow the
information in the game tree and from that point on has a winning
strategy. Thus, what the program playing as the first player could do
is to maximize the probability that the second player makes a mistake.

At the beginning, all moves lead to R-LOSS successors. However, some of
these successors may themselves have R-WIN successors that could be
chosen by Green carelessly in the next move. Red can look one step ahead
and choose a move that leads to a position having more R-WIN successors
than the others. Then, for a human playing as Green, it might be more
difficult not to choose an R-WIN successor at the next step, thus leading
to the program's victory.

Let us assume Red is using this method to make his choice. He
definitely has a higher chance to win than choosing blindly by just
avoiding triangles, under the condition that Green is not perfect and
sometimes chooses his moves randomly. Is there a way to make it even
easier for Green to make a mistake? One way could be to take into
account small perceptive preferences of the human higher visual
system. Empirically, the six edges in the square at the center of the
hexagon seem to be slightly more eye-catching than the other four
side-edges, and these seem easier to perceive than the remaining five
edges. Apart from the quantitative value (the number of R-WIN
successors), each move thus has its qualitative value (whether the R-WIN
successor is easily reached), the latter depending on the current game
layout. Red has a good chance of winning against a human playing as
Green by selecting with higher probability a move that scores higher
in this combined heuristic function. Next, we explain how this
function is further refined through learning to take counter-measures
against the learning capabilities of human players.

\subsection*{Learning}

We now have a heuristic strategy for Red which works as follows: The
program will choose its best move by considering both the number of
R-WIN positions following this move and whether such R-WIN positions
are likely to be chosen by Green. This approach sounds reasonable, but
once a human player has found a sequence of moves leading to his
victory, he can always replay this sequence and win again, assuming
that he can restart the game in case the program randomly chooses a
move that does not follow his sequence. As a counter-strategy, the
first player was made able to learn from experience.

Whenever the program played as the first player, Red, it analyzes its
moves at the end of the game. If it won by forcing the human to give
up, that is, the human did not lose by closing a triangle merely by
mistake, without being forced to do so, the program's moves were
apparently well chosen and the sequence should be memorized and chosen
again with increased likelihood. With this knowledge, the program will
tend to behave identically if the same or an isomorphic situation is
encountered in the future. On the other hand, if the program has lost
the game, it will know that its moves may not have been good enough
ones, and thus will decrease the probability of selecting such moves
or isomorphic ones in later plays.

All the R-WIN/R-LOSS information for each move is stored in a position
table. Each entry of the table is a number representing an R-WIN or
R-LOSS. All R-WIN entries have the value 0 because no additional
information is necessary: Once reached, Red's strategy is to always
select an R-WIN move. The value of an R-LOSS entry changes according to
its learned desirability for Red from $-128$ to $127$, the range of a
byte in Java. Each entry is initially set to 1 to distinguish it from
R-WIN entries. This value will be modified according to the analysis
made after each play. If the program has won, the value of the
corresponding R-LOSS entries will be increased by a learning factor,
and decreased if it lost, with a maximum and minimum value.
Additionally, value 0 is always avoided when learning as it is
reserved for R-WIN. The values held by R-LOSS entries are used during
heuristic play to probabilistically guide the selection of the moves
by the computer, together with the static heuristic function already
described. At the very beginning, no experience is available, so the
selection of moves is only based on the static heuristic function. As
more and more experience accumulates, the learned information becomes
the dominant factor in the selection of moves. Since changes are
memorized in the compressed position table, learning is done very
efficiently for a multitude of isomorphic game sequences.

Because the program is available in form of a Java applet, it can
easily recontact the Internet server from whom it was originally
downloaded. Through a pre-specified Internet port, it hands back the
acquired playing experience. On the server side, a Java daemon is
running which registers all information communicated through Java
applets on client sides. If no contact can be established because the
Internet connection is down, the information is lost. However, through
the same client-server connection, human players who have won are
allowed to enter their nickname into a hall-of-fame stored at the
server-side, which may be an incentive to allow the Internet
connection to be reestablished at the end of the play (relevant for
non-permanent Internet access only). To make it more interesting, this
hall-of-fame is sorted according to the amount of time that was needed
to win against the program. In order to avoid that a clever person
places his or her nickname on top of the list by quickly letting two
Java applets play against each other, entering one's nickname is only
possible in the ``Allow shaking'' mode: In this mode, an animated
random permutation of the six board vertices is performed after each
move of the program, thus resulting in an isomorphic game situation
which nevertheless looks quite different to the human eye and makes it
more difficult to let two applets play against each other. Of course,
these permutations have to be taken already into account when
selecting the next move according to the static heuristic function
described in the previous section.

The combination of perfect information with the static heuristic
function alone makes the program already very strong. The learning
additionally ensures that it will be unlikely that someone will win
repeatedly against the program. According to our experience, it is now
counter-intuitively hard to win against the program, and even harder to
win again. We therefore assert:

\begin{claim}
Following the described adaptive strategy together
with our complete game tree data means that we have come as close as
possible to\/ {\em ultra-strongly solving} Sim in the sense of
Definition~\ref{def:solved}.
\end{claim}

\subsection{Ultra-strongly solving Sim$^+$}

What happens if we allow that each player colors more than one edge
during his turn to move? This variant which we called \gramseyplus\
intuitively corresponds even closer to the results from Ramsey theory
than \gramsey\ does, since the relation between red and green edges
can vary arbitrarily. Obviously, this game is more difficult to
analyze as we are confronted with a larger game tree. For Sim$^+$, the
first move no longer consists in choosing one edge from fifteen but an
arbitrary selection out of the 15 possible edges.

In the original Sim game, it is clear at each step who should move
next because each player colors only one edge at a time. Thus, one
table is enough for both players since it is always clear from the
number of edges which player's turn it is. In this new variant,
however, it is impossible to tell which one should move next merely
from the game position. We have to use two position tables to store
the strategies of the first and second player. Only a few
modifications were needed to extend Sim to allow more edges to be
colored at each player's turn. The same methods were used to generate
the whole game tree (the directed acyclic game graph has 13158 entries
and thus is more than five times as large as in the standard variant
Sim), classify the positions and then save them in two tables with
different R-WIN/R-LOSS information. This game variant is available
from the same Internet address by choosing option ``Allow more moves
each time'' in the applet's control panel.

\begin{theo}\label{theo:simplusloss}
The second player, Green, has a winning strategy in Sim$^+$ =
\gramseyplus$(K_{\scriptsize \ramsey(3,3)},K_3,\{\},\{\})$.
\end{theo}

\noindent As with Theorem~\ref{theo:simloss}, the proof of this
statement is more or less a very long enumeration of cases which we do
not include here. It can easily be reconstructed from the data
available together with the Java-applet on the author's home-page.

Accordingly, it is again the second player who has a winning strategy.
In our Java applet, we implemented both this winning strategy and a
heuristic learning counter-strategy almost identical to the one for
Sim, thus allowing us again to assert the following:

\begin{claim}
We have come as close as possible to\/ {\em ultra-strongly solving} the
game \gramseyplus$(K_{\scriptsize \ramsey(3,3)},K_3,\{\},\{\})$.
\end{claim}

\subsection{Sim$_4$ and above}\label{sec:sim4}

Another challenge may be a similar game based on \ramsey$(4,4)$. With
this variant, we need \ramsey$(4,4)=18$ (\citet{Greenwood55}) vertices
with ${18\choose2}=153$ edges between them to play a game analogous to
Sim. Since the structure to avoid would be $K_4$, i.e., a tetrahedron
(a pyramid with four triangular faces), one could play this game in
simulated three-dimensional space. The crucial question is again the
number of possible non-isomorphic game positions. The exact number can
be bounded from above by counting the number of non-isomorphic game
positions including non-legal positions containing several
monochromatic tetrahedra. This task can be solved by straightforward
application of P\'olya's Theorem \cite{Polya37} (or `P\'olya's
enumeration formula') on counting orbits under group actions using
Harary's cycle index for the group $S_{18}^2$ of edge permutations of
$K_{18}$ \cite{Harary55} which enables us to count colorings which are
distinct with respect to the action of above permutation group (for a
survey see \citet{Harary73}, for a gentle introduction
\citet[Chap.~9]{Tucker95}; the following special case is proved and
explained in detail in \citet{Gessel95}):
\begin{eqnarray*}
Z(S_n^2) & = & \sum_{\scriptsize \begin{array}{c}
                     (m_1,m_2,\ldots,m_n)\in [1,n]^n\\
                     \sum_{i=1}^n i m_i=n
                   \end{array}} \frac{1}{\prod_{k=1}^n (k^{m_k} m_k!)}
             \prod_{k=1}^{\lfloor n/2\rfloor}(p_k p_{2k}^{k - 1})^{m_{2k}} \times\\
         &&  \prod_{k=1}^{\lfloor (n-1)/2\rfloor}p_{2k+1}^{km_{2k+1}} 
             \prod_{k=1}^n p_k^{k{m_k\choose 2}} 
             \prod_{\scriptsize \begin{array}{c}
                     (i,j)\in [1,n]^2\\
                     i<j
                   \end{array}}
                p_{\lcm(i,j)}^{\gcd(i,j)\;m_i m_j}
\end{eqnarray*}
where lcm and gcd denote the least common multiple and greatest common
divisor. The number of non-isomorphic edge-red-green-colorings of
$K_{18}$ with $r$ red and $g$ green edges is the coefficient of
$x^ry^g$ in $Z(S_{18}^2)$ when we replace each $p_i$ with $1+x^i+y^i$, denoted by 
\begin{eqnarray*}
|K^{(r,g)}_{18}|&\de&[x^ry^g]\;Z(S_{18}^2)\;\rule[-1.5ex]{.3pt}{3.5ex}\;\raisebox{-1ex}{\scriptsize $p_i \:\rightarrow\: 1+x^i+y^i$}
\end{eqnarray*}
Summing up over all legal $(r,g)$ tuples, we found that the number
of non-isomorphic game positions is thus bounded by a number larger
than $10^{56}$\renewcommand{\thefootnote}{\ddag}~\footnote[1]{The
exact value, computed with P\'olya's enumeration formula from above,
of partial edge-red-green-colorings of $K_{18}$ such that the number
of red edges is always equal or one larger than the number of green
edges is
\[122817954504260150325481627994395745196940238595512818831.\]}\renewcommand{\thefootnote}{\arabic{footnote}}.

A fifty times smaller number, $2\times 10^{54}$, came out by
probabilistically counting only legal non-isomorphic game positions
through sampling using Monte Carlo methods. The corresponding
experiments were conducted in two ways, using two methodically
completely independent ways to sample. As one sees below, the
confidence intervals of the two methods overlapped, boosting our trust
in the results:

\begin{enumerate}
\item method, where the mean is computed by the following formula:
\begin{eqnarray*}
L_1(18,4)&\de&\sum_{\scriptsize \begin{array}{c}(r,g)\in \N^2\\r=g \mbox{~~or~~} 
r=g+1\\ r+g\leq {18\choose 2}\end{array}} \frac{1}{M}\: \sum_{k=1}^M\:
\left(\frac{{{18\choose 2}\choose r}{{18\choose 2} - r\choose g}}{E(\mathcal{G}_k(r,g))}
\: T_4(\mathcal{G}_k(r,g))\right)
\end{eqnarray*}
where $M=1000$ was the constant
sampling size for each $(r,g)$ class, $\mathcal{G}_k(r,g)$ being the
$k$-th randomly and uniformly drawn edge-red-green-coloring of
$K_{18}$ with $r$ red edges and $g$ green edges,
$E(\mathcal{G}_k(r,g))$ the size of the isomorphism class of
$\mathcal{G}_k(r,g)$ with respect to edge permutations, and
$T_n(\mathcal{G}_k(r,g))$ being defined as zero if $\mathcal{G}_k(r,g)$
contained a monochromatic $K_n$, and one otherwise. By the
formula, we adjust the ratio of graphs found to contain no
monochromatic tetrahedron when sampling 1000 times over each class by
multiplying with the complete population and dividing by the size of
the isomorphism class, thus taking into account the various sizes of
isomorphism classes and correcting the sampling error, and then sum up
over all legal $(r,n)$ tuples. The resulting .99
confidence interval, with the mean value inserted in the middle, is:

\[[1.7\times 10^{54}\setcomma 2.2\times 10^{54}\setcomma 2.7\times 10^{54}]\]

\item method, using P\'olya's enumeration formula from above to
compute the number of non-isomorphic graphs for an $(r,g)$ coloring
with the same restrictions as above on $(r,g)$, this time using
different sampling sizes for different $(r,g)$ values, using the
following formula for the mean value:
\begin{eqnarray*}
L_2(18,4)&\de&\sum_{\scriptsize \begin{array}{c}(r,g)\in \N^2\\r=g \mbox{~~or~~} 
r=g+1\\ r+g\leq {18\choose 2}\end{array}} \frac{\left(\sum_{k=1}^{M_{r+g}}\:
T_4(\mathcal{G}_k(r,g))\right)}{M_{r+g}}\: |K^{(r,g)}_{18}|
\end{eqnarray*}
where the sampling rate $M_{r+g}$ was 100 for low $(r,g)$ tuples and
increased up to 50000 for larger values of $(r,g)$, the other variables
and functions being defined as before. This formula computes the ratio
of valid colorings that do not contain monochromatic tetrahedra,
multiply it with the number of non-isomorphic $(r,g)$ colorings of
$K_{18}$, and then again sum up over all legal $(r,n)$ tuples. The
resulting .99 confidence interval, with the mean value inserted in the
middle, is:

\[[1.9\times 10^{53}\setcomma 2.0\times 10^{54}\setcomma 5.0\times 10^{54}]\]
\end{enumerate}

\noindent We thus can safely assume that the number $2\times 10^{54}$
is not off by too many orders of magnitude compared to the real size
of the directed acyclic game graph. Generating a graph of this size
would require around $10^{21}$ centuries even if 300 trillion nodes
could be generated during each second by each one of one billion
computers running in parallel all the time, sloppily assuming for the
sake of the argument that the computation lends itself to such massive
parallelization. To put this time span into perspective, this is ten
trillion times the currently estimated amount of time \cite{Krauss98}
since the Big Bang.

\begin{claim}\rem{genauer, mit \"Ubersichtstable und allen Werten, 
cf \citet{Allis94}}
In view of above numbers and the \pspacec{}ness of the general
\gramsey\ game, we claim that to compute a winning strategy for
$\gramsey(K_{\scriptsize \ramsey(n,n)},K_n,\{\},\{\})$ (as well as for the
$\gramseyplus(K_{\scriptsize \ramsey(n,n)},K_n,\{\},\{\})$ variant) is, from all
practical points of view, intractable for $n>3$.
\end{claim}

\noindent These result deterred us from attempting to compute the
compressed game tree and thus the winning strategy for the bigger
variants of Sim, but a heuristic strategy might still be of
interest. As mentioned already earlier, Ramsey numbers $(n,n)$ with
$n$ greater than four are open research problems that recently
generated a lot of interest~\cite{Radziszowski94}, and plausibly lead
to more and more complicated games.

\section{Further observations}\label{sec:open}

In the previous section, we showed that it is very likely that we
never will be able to weakly solve $\gramsey(K_{\scriptsize
\ramsey(n,n)},K_n,\{\},\{\})$ for $n>3$. However, it might be possible
to at least ultra-weakly solve these games, as described below for the
\gramsey$(K_{18},K_4,\{\},\{\})$ case:

From \ramsey$(4,4)=18$, we know that all
\[
|K^{(77,76)}_{18}|\: = \: 114722035311851620271616102401\: >\: 10^{29}
\]
non-isomorphic $(r,g)=(77,76)$ edge-red-green-colorings of $K_{18}$
contain at least one monochromatic tetrahedron. We also know that
there exists one $(76,76)$ edge-red-green-coloring of $K_{18}$ (one
edge remaining uncolored) that contains no monochromatic tetrahedron,
through the following fact:

\begin{fact}[adapted from Staszek Radziszowski]\label{fact:staszek} \
There is a unique edge-red-green-coloring of $K_{17}$ without monochromatic
tetrahedron, call it $C$, where the number of edges of the same color
leaving any vertex is equal to 8. Take any vertex $v$ of $C$, make
its duplicate $u$, i.e., edges $\{x\setcomma v\}$ and $\{x\setcomma
u\}$ have the same color, for all $x$. $C$ extended by $u$ is the
desired $(76,76)$ edge-red-green-coloring of $K_{18}$ containing no
monochromatic tetrahedron, where edge $\{u\setcomma v\}$ is not
colored, and it is unique up to isomorphism. 
\end{fact}

\noindent Because of above observations and the fact that there are
${18\choose 2}=153$ edges, that is, an odd number, that can be colored
altogether, a natural question would be whether the second player has
a winning strategy in \gramsey$(K_{18},K_4,\{\},\{\})$.  This question
as well as Fact~\ref{fact:staszek} (without the uniqueness, and by
being careful in the choice of the vertex to be duplicated) can easily
be generalized to larger symmetric binary Ramsey numbers:

\begin{open}\label{open:even}
Consider \gramsey$(K_k,K_n,\{\},\{\})$ where $k=\ramsey(n,n)$. Is it
always true that the first player has a winning strategy in this game
iff\/ ${k\choose 2}$ is even?
\end{open}

\noindent If we could show that a player with a winning strategy in
some small game instance of \gramsey$(K_k,K_n,\{\},\{\})$ where
$k=\ramsey(n,n)$, such as Sim, can always force a win already
{\em before\/} move ${k\choose 2}$, this obviously would strongly
indicate that the answer to the question in Open
Problem~\ref{open:even} is no.  We tested in on Sim's directed acyclic
game graph but found that the second player cannot force a win before
move 15, so the problem remains open.  Open Problem~\ref{open:even} can
be generalized to \gramsey$(G,A,E^r,E^g)$ games as follows: 

\begin{open}\label{open:geven}
Consider\/ \gramsey$(G,A,E^r,E^g)$, where
\begin{eqnarray*}
c & \de & \min_{\scriptsize \makebox[14ex]{$\begin{array}{c}(r,g)\in \N^2\\r=g \mbox{~~or~~} 
r=g+1\\ r+g\leq {|E(G)|-|E^r|-|E^g|}\end{array}$}}\{r+g \: | \: (G,E^r,E^g)^{(r,g)}\rightarrow A\}\;,
\end{eqnarray*}
and where $(G,E^r,E^g)^{(r,g)}$ denotes an $(r,g)$
edge-red-green-coloring of the uncolored edges of the precolored graph
$(G,E^r,E^g)$.  Is it always true that the first player has a winning
strategy in this game iff $c$ is even?
\end{open}

\noindent It is, however, rather unlikely that Open
Problem~\ref{open:geven} has a positive answer since together with
Theorem~\ref{theo:piptc} and Theorem~\ref{theo:gramsey}, this would
imply that \pspace = \pipt\ and that the polynomial hierarchy
collapses to its second level, which would be very surprising.

Regarding our asymptotic results, we could prove that all unrestricted
graph Ramsey games are \pspacec.  However, a word of caution might be
appropriate. \pspacec{}ness implies that, under the condition that
\p$\,\not=\,$\pspace, there exists no efficient algorithm to decide
whether a {\em general\/} position allows a forced win for the first
player. Unfortunately, such a proof says nothing about the most
important position of all, namely the uncolored graph. It could be
that there is a simple winning strategy for player Red given an
uncolored graph, and all situations proved hard never occur during
optimal play. This leads us to the following conjecture:

\begin{conj}\label{conj:empty} 
Graph Ramsey games played on $(G,A,\{\},\{\})$ are \pspacec.
\end{conj}

\noindent However, given the evidence of tractable subcases (cf.\
Theorem~\ref{theo:lseekramsey}) we also believe that:

\begin{conj} Graph Ramsey achievement games played on $(K_n,A,E^r,E^g)$
are tractable.
\end{conj}

\noindent We still believe that graph Ramsey avoidance games restricted
to graphs based on classic symmetric binary Ramsey numbers are
difficult problems:

\begin{conj}\label{conj:classic}
Graph Ramsey avoidance games played on $(K_k,K_n,E^r,E^g)\,$ where
$\:k\geq\ramsey(n,n)\:$ are \pspacec.
\end{conj}

\noindent In contrast, the complexity of avoidance games based on
\ramsey\ numbers where we specify only the size $n$ of the
monochromatic complete graph $A=K_n$ is conjectured to lie well beyond
\pspace, since it requires the computation of explicit Ramsey numbers
(for exponential lower {\em and\/} upper bounds of classic symmetric
binary Ramsey numbers, see \cite{Graham90,Nesetril95}; as mentioned,
only the first two nontrivial of these numbers are known so far
\cite{Radziszowski94}) and manipulations on graphs of the size of
these numbers, thus suggesting doubly exponential space requirements
because of the succinct input representation. The combination of
Conjectures~\ref{conj:empty} and \ref{conj:classic} then
leads~to\hidden{}{tilde}:

\begin{conj}\label{conj:texpspace} The graph Ramsey avoidance games played on 
$(K_{\scriptsize {\ramsey(n,n)}},K_n,\{\},\{\})$ are
2-\expspace-complete.
\end{conj}

\noindent Note that if Open Problem~\ref{open:even} is answered
positively, the complexity in Conjecture~\ref{conj:texpspace} would
have to be changed to \expspace-complete.

We could not show meaningful restrictions on the achievement
graph $A$ as in Corollaries~\ref{cor:avoidrestriction} and
\ref{cor:achieverestriction} for the \gachieveramsey\ game. We also
could not find any meaningful restriction on the game graphs $G$ for
any of the graph Ramsey games, even though the construction in the
proof of Theorem~\ref{theo:gweakachieveramsey} looks promising:
Despite the easily proved fact that \QBF(3CNF) is \pspacec\ even if
each variable occurs less than 6 times in the propositional formula,
the construction used by T.~\citet{Schaefer78} to prove the
\pspacec{}ness of \gposdnf\ adds variables occurring linearly in the
number of clauses of the original 3CNF formula, and there seems to be
no way to get rid of these occurrences since only positive literals
are allowed in \gposdnf.

\begin{open}
Show that \gachieveramsey\ remains \pspacec\ even if the achievement
graph $A$ is restricted to a meaningful subclass of graphs such as
fixed, bipartite or degree-restricted graphs.
\end{open}

\begin{open}
Show that Theorems~\ref{theo:gramsey}--\ref{theo:gachieveramsey} hold
even if the game graph $G$ is restricted to a meaningful subclass of
graphs such as bipartite or degree-restricted graphs.
\end{open}

\noindent Other interesting directions of research include graph
Ramsey games played on directed graphs.  Plausibly, they will be as
difficult as their undirected versions, but might be useful for the
analysis of different real world applications. Also, transfinite graph
Ramsey avoidance games in the spirit of
\cite{Beck81,Beck83,Hajnal84,Knor96,Komjath84,Pekec96} where players
must color $\aleph_0$ many edges per move are conceivable, their
decision problems likely being questions of computability rather than
of complexity.

\section{Concluding remarks}

Sim and Sim$^+$ are very easy to learn and can be played on a small
piece of paper, a typical game taking only a few
minutes. Nevertheless, they are fascinating to play because they are much
more difficult than it first appears while at the same time being simple
and elegant. In this paper, we proved that these games belong to a
family of graph Ramsey games that are \pspacec, implying that they are
the most difficult problems in a class of problems generally believed
to be intractable, though a formal proof that $\p\not=\pspace$ is
lacking. At the very least, our results imply that the studied games
are equivalent from the point of view of structural complexity theory
to a large number of well-known games (e.g., Go~\cite{Lichtenstein80})
and problems of industrial relevance (e.g., decision-making under
uncertainty such as stochastic scheduling~\cite{Papadimitriou85})
generally recognized as very difficult. The new characterization of
\pspacec\ problems as graph Ramsey games might help in studying
competitive situations from industry, economics or politics where
opposing parties try to achieve or to avoid a certain pattern in the
structure of their commitments, in particular situations that may
arise in distributed networks, maybe in a future not too far away
(cf.\ for instance the mobile Internet agent warfare scenarios
described in \cite{Thomsen98}).

We also explained how we constructed a perfect second and an
ultra-strong heuristic first player for Sim and Sim$^+$ that can now
be played with an attractive graphical interface on any computer for
which a Java compatible browser is available.  Our program is able to
learn persistently from past experience by playing with different
persons through the Internet. Similar self-improving techniques with
client-server style learning over the Internet could be applied to
other games, but also to more down-to-earth applications such as
tutoring systems, intelligent language tools such as intelligent
dictionaries or grammar and style checkers, intelligent agents, or
distributed manufacturing systems.
 
Additionally, we showed that it is highly unlikely that similar games
based on symmetric binary Ramsey numbers for $n>3$ will ever be even
weakly solved. Finally, we listed a number of open problems and
conjectures related to graph Ramsey games.

\section*{Acknowledgments}

I am very indebted to my students Andreas Beer and Bidan Zhu for doing
the excellent implementation work which resulted in the Java applet
and in Zhu's Master's thesis~\cite{Zhu97}. Burki Zimmermann kindly
helped me with P\'{o}lya's enumeration formula, Fritz Leisch and Jack
Tomsky with binomial and poisson distributions to compute the
confidence intervals (all used in Section~\ref{sec:sim4}). I would
like to thank them, Thomas Eiter, Uri Feige, Aviezri Fraenkel, Georg
Gottlob, Frank Harary, Staszek Radziszowski, G\"{u}nter Schachner,
Helmut Veith, Oleg Verbitzky, as well as Mutsunori Yagiura for helpful
suggestions and stimulating discussions. This research was partially
supported by Austrian Science Fund Project N Z29-INF as well as by a
research fellowship from the Japan Society for the Promotion of
Sciences, the latter allowing me to work for two months at the
department of Toshihide Ibaraki at Kyoto University in 1997. I am also
very grateful to ``Jimmy'' Werner DePauli-Schimanovich-G\"{o}ttig for
urging me to implement the perfect and learning player~\cite{Slany88},
to Ranan Banerji for introducing me to Ramsey theory, and to my
classmates from high-school who competed with me in programming
heuristic players for Sim on our various pocket calculators.

\bibliographystyle{authordate1sml} 
\bibliography{../ramsey}

\begin{thebibliography}{}
\setlength{\itemsep}{1pt plus 2pt}
\small

\bibitem[\protect\citename{Allis, }1994]{Allis94}
Allis, Louis~Victor. 1994.
\newblock {\em Searching for solutions in games and artificial intelligence}.
\newblock Ph.D. thesis, University of Limburg.
\newblock ftp://ftp.cs.vu.nl/pub/victor\linebreak[0]/PhDthesis/thesis.ps.Z.

\bibitem[\protect\citename{Beck \& Csirmaz, }1982]{Beck82}
Beck, J. \& Csirmaz, L. 1982.
\newblock Variations on a game.
\newblock {\em Journal of Combinatorial Theory, Series A}, {\bf 33}, 297--315.

\bibitem[\protect\citename{Beck, }1981]{Beck81}
Beck, J\'ozsef. 1981.
\newblock {Van der Waerden} and {Ramsey} type games.
\newblock {\em Combinatorica}, {\bf 1}(2), 103--116.

\bibitem[\protect\citename{Beck, }1983]{Beck83}
Beck, J\'ozsef. 1983.
\newblock Biased {Ramsey} type games.
\newblock {\em Studia Scientiarum Mathematicarum Hungarica}, {\bf 18},
  287--292.

\bibitem[\protect\citename{Beck, }1997]{Beck97}
Beck, J\'ozsef. 1997.
\newblock Graph games.
\newblock {\em In:} {\em Proceedings of the International Colloquium on
  Extremal Graph Theory}.

\bibitem[\protect\citename{Burr, }1987]{Burr87}
Burr, Stefan~A. 1987.
\newblock What can we hope to accomplish in generalized {R}amsey theory?
\newblock {\em Discrete Mathematics}, {\bf 67}, 215--225.

\bibitem[\protect\citename{Cameron, }1994]{Cameron94}
Cameron, Peter~J. 1994.
\newblock {\em Combinatorics: topics, techniques, algorithms}.
\newblock Cambridge University Press.

\bibitem[\protect\citename{Chv\'atal \& Harary, }1972]{Chvatal72}
Chv\'atal, V. \& Harary, Frank. 1972.
\newblock Generalized {R}amsey theory for graphs.
\newblock {\em Bull.\ Amer.\ Math.\ Soc.}, {\bf 78}, 423--426.

\bibitem[\protect\citename{Cook \& Shader, }1979]{Cook79}
Cook, Michael~L. \& Shader, Leslie~E. 1979.
\newblock A strategy for the {R}amsey game ``{T}ritip''.
\newblock {\em Pages  315--324 of:} {\em Proc. 10th southeast. Conf.
  Combinatorics, Graph Theory and Computing, Boca Raton}.
\newblock Congr. Numerantium 23, vol. 1.

\bibitem[\protect\citename{Cornelius, }1991]{Cornelius91}
Cornelius, Michael. 1991.
\newblock {\em What's your game?}
\newblock Cambridge University Press.

\bibitem[\protect\citename{Darby \& Laver, }1998]{Darby98}
Darby, Carl \& Laver, Richard. 1998.
\newblock Countable length {R}amsey games.
\newblock {\em Pages  41--46 of:} et~al., Carlos Augusto~{Di Prisco} (ed), {\em
  Set theory: techniques and applications. Proceedings of the conferences,
  Curacao, Netherlands Antilles, June 26--30, 1995 and Barcelona, Spain, June
  10--14, 1996}.
\newblock Dordrecht: Kluwer Academic Publishers.

\bibitem[\protect\citename{DeLoach, }1971]{DeLoach71}
DeLoach, A.~P. 1971.
\newblock Some investigations into the game of {SIM}.
\newblock {\em J. Recreational Mathematics}, {\bf 4}(1), 36--41.

\bibitem[\protect\citename{Deuber, }1975]{Deuber75}
Deuber, W. 1975.
\newblock A generalization of {R}amsey's theorem.
\newblock {\em Pages  323--332 of:} Hajnal, A.; Rad\'o, R.; \& S\'os, V.T.
  (eds), {\em Infinite and finite sets}.
\newblock Colloq. Math. Soc. J\'anos Bolyai, vol. 10.
\newblock North-Holland.

\bibitem[\protect\citename{Diestel, }1997]{Diestel97}
Diestel, Reinhard. 1997.
\newblock {\em Graph theory}.
\newblock Graduate Texts in Mathematics, vol. 173.
\newblock Springer.

\bibitem[\protect\citename{Engel, }1972]{Engel72}
Engel, Douglas. 1972.
\newblock {DIM}: three-dimensional {S}im.
\newblock {\em J. Recreational Mathematics}, {\bf 5}, 274--275.

\bibitem[\protect\citename{Erd\H{o}s \& Selfridge, }1973]{Erdos73}
Erd\H{o}s, P. \& Selfridge, J.~L. 1973.
\newblock On a combinatorial game.
\newblock {\em Journal of Combinatorial Theory, Series A}, {\bf 14}, 298--301.
\newblock (The self-reference on the paper falsely claims it is in Series B).

\bibitem[\protect\citename{Erd\H{o}s \& Szekeres, }1935]{Erdos35}
Erd\H{o}s, P. \& Szekeres, D. 1935.
\newblock A combinatorial problem in geometry.
\newblock {\em Compos. Math.}, {\bf 2}, 463--470.

\bibitem[\protect\citename{Erd\H{o}s {\em et~al.}, }1975]{Erdos75}
Erd\H{o}s, Paul; Hajnal, Andr\'as; \& P\'osa, Latjos. 1975.
\newblock Strong embeddings of graphs into colored graphs.
\newblock {\em Pages  585--595 of:} Hajnal, A.; Rad\'o, R.; \& S\'os, V.T.
  (eds), {\em Infinite and finite sets}.
\newblock Colloq. Math. Soc. J\'anos Bolyai, vol. 10.
\newblock North-Holland.

\bibitem[\protect\citename{Even \& Tarjan, }1976]{Even76}
Even, S. \& Tarjan, R.~E. 1976.
\newblock A combinatorial problem which is complete in polynomial space.
\newblock {\em Journal of the ACM}, {\bf 23}, 710--719.
\newblock Also appeared in Proc. 7th Ann. ACM Symp. Theory of Computing
  (Albuquerque, NM, 1975), Assoc. Comput. Mach., New York, NY, 1975, pp.
  66--71.

\bibitem[\protect\citename{Exoo, }1980-81]{Exoo80}
Exoo, Geoffrey. 1980-81.
\newblock A new way to play {R}amsey games.
\newblock {\em J. Recreational Mathematics}, {\bf 13}(2), 111--113.

\bibitem[\protect\citename{Fraenkel \& Lichtenstein, }1981]{Fraenkel81}
Fraenkel, A.~S. \& Lichtenstein, D. 1981.
\newblock Computing a perfect strategy for $n\times n$ {C}hess requires time
  exponential in $n$.
\newblock {\em Journal of Combinatorial Theory, Series A}, {\bf 31}, 199--214.
\newblock Preliminary version in Proc. 8th Internat. Colloq. Automata,
  Languages and Programming (S. Even and O. Kariv, eds.), Acre, Israel, 1981,
  {\it Lecture Notes in Computer Science\/} {\bf 115}, 278--293, Springer
  Verlag, Berlin.

\bibitem[\protect\citename{Fraenkel {\em et~al.}, }1978]{Fraenkel78}
Fraenkel, A.~S.; Garey, M.~R.; Johnson, D.~S.; Schaefer, T.; \& Yesha, Y. 1978
  (Oct.).
\newblock The complexity of {C}heckers on an $n\times n$ board --- preliminary
  report.
\newblock {\em Pages  55--64 of:} {\em Proc. 19th Ann. Symp. Foundations of
  Computer Science}.
\newblock IEEE Computer Soc., Long Beach, CA, Ann Arbor, MI.

\bibitem[\protect\citename{Fraenkel, }1991]{Fraenkel91}
Fraenkel, Aviezri~S. 1991.
\newblock Complexity of games.
\newblock {\em Pages  111--153 of:} Guy, Richard~K. (ed), {\em Combinatorial
  Games}.
\newblock Proceedings of Symposia in Applied Mathematics, vol. 43.
\newblock American Mathematical Society.

\bibitem[\protect\citename{Fraenkel, }1994 (revised October 7,
  1999)]{Fraenkel94}
Fraenkel, Aviezri~S. 1994 (revised October 7, 1999).
\newblock Dynamic surveys in combinatorics: Combinatorial games: Selected
  bibliography with a succinct gourmet introduction.
\newblock {\em Electronic Journal of Combinatorics}.
\newblock http://www.combinatorics.org/Surveys/ds2.ps.

\bibitem[\protect\citename{Funkenbusch, }1971]{Funkenbusch71}
Funkenbusch, W.~W. 1971.
\newblock {SIM} as a game of chance.
\newblock {\em J. Recreational Mathematics}, {\bf 4}(4), 297--298.

\bibitem[\protect\citename{Gardner, }1973a]{Gardner73a}
Gardner, Martin. 1973a.
\newblock Mathematical games.
\newblock {\em Scientific American}, {\bf 228}(1), 108--115.

\bibitem[\protect\citename{Gardner, }1973b]{Gardner73b}
Gardner, Martin. 1973b.
\newblock Mathematical games.
\newblock {\em Scientific American}, {\bf 228}(5), 102--107.

\bibitem[\protect\citename{Gardner, }1986]{Gardner86}
Gardner, Martin. 1986.
\newblock {\em Knotted doughnuts and other mathematical entertainments}.
\newblock W. H. Freeman and Company.
\newblock Chap.~9, pages  109--122.

\bibitem[\protect\citename{Garey \& Johnson, }1979]{Garey79}
Garey, Michael~R. \& Johnson, David~S. 1979.
\newblock {\em Computers and intractability: A guide to the theory of
  {NP}-completeness}.
\newblock Freeman and Co.

\bibitem[\protect\citename{Gessel \& Stanley, }1995]{Gessel95}
Gessel, Ira~M. \& Stanley, Richard~P. 1995.
\newblock Algebraic Enumeration.
\newblock {\em Chap. 21, pages  1021--1061 of:} Graham, R.; Gr\"otschel, M.; \&
  Lov\'asz, L. (eds), {\em Handbook of Combinatorics},  vol. II.
\newblock Elsevier Science.

\bibitem[\protect\citename{Graham {\em et~al.}, }1990]{Graham90}
Graham, Ronald~L.; Rothschild, Bruce~L.; \& Spencer, Joel~H. 1990.
\newblock {\em Ramsey theory}. 2nd edn.
\newblock Wiley.

\bibitem[\protect\citename{Greenwood \& Gleason, }1955]{Greenwood55}
Greenwood, R.~E. \& Gleason, A.~M. 1955.
\newblock Combinatorial relations and chromatic graphs.
\newblock {\em Canadian Journal of Mathematics}, {\bf VII}, 1--7.

\bibitem[\protect\citename{Guy, }1991]{Guy91}
Guy, Richard~K. 1991.
\newblock What is a game?
\newblock {\em Pages  1--21 of:} Guy, Richard~K. (ed), {\em Combinatorial
  Games}.
\newblock Proceedings of Symposia in Applied Mathematics, vol. 43.
\newblock American Mathematical Society.

\bibitem[\protect\citename{Hajnal \& Nagy, }1984]{Hajnal84}
Hajnal, A. \& Nagy, Zs. 1984.
\newblock Ramsey games.
\newblock {\em Transactions of the {A}merican Mathematical Society}, {\bf
  284}(2), 815--827.

\bibitem[\protect\citename{Harary \& Palmer, }1973]{Harary73}
Harary, F. \& Palmer, E. 1973.
\newblock {\em Graphical Enumeration}.
\newblock New York: Academic Press.

\bibitem[\protect\citename{Harary, }1955]{Harary55}
Harary, Frank. 1955.
\newblock The number of linear, directed, rooted and connected graphs.
\newblock {\em Trans.\ Am.\ Math.\ Soc.}, {\bf 78}, 445--463.

\bibitem[\protect\citename{Harary, }1962]{Harary62}
Harary, Frank. 1962.
\newblock Personal communication.

\bibitem[\protect\citename{Harary, }1982]{Harary82}
Harary, Frank. 1982.
\newblock Achievement and avoidance games for graphs.
\newblock {\em Pages  111--119 of:} Bollob\'as, B\'ela (ed), {\em Proceedings
  of the Conference on Graph Theory}.
\newblock Ann. Discrete Math., vol. 13.
\newblock Cambridge: North-Holland, mathematics studies 62.

\bibitem[\protect\citename{Knor, }1996]{Knor96}
Knor, Martin. 1996.
\newblock On {R}amsey-type games for graphs.
\newblock {\em Australasian J. Comb.}, {\bf 14}, 199--206.

\bibitem[\protect\citename{Komj\'ath, }1984]{Komjath84}
Komj\'ath, P\'eter. 1984.
\newblock A simple strategy for the {R}amsey-game.
\newblock {\em Studia Scientiarum Mathematicarum Hungarica}, {\bf 19},
  231--232.

\bibitem[\protect\citename{Krauss, }1998]{Krauss98}
Krauss, Lawrence~M. 1998.
\newblock The End of the Age Problem, and the Case for a Cosmological Constant
  Revisited.
\newblock {\em Astrophysical Journal}, {\bf 501}(2), 461--466.

\bibitem[\protect\citename{Lichtenstein \& Sipser, }1980]{Lichtenstein80}
Lichtenstein, D. \& Sipser, M. 1980.
\newblock Go is polynomial-space hard.
\newblock {\em J. Assoc. Comput. Mach.}, {\bf 27}, 393--401.
\newblock Earlier draft appeared in Proc. 19th Ann. Symp. Foundations of
  Computer Science (Ann Arbor, MI, Oct. 1978), IEEE Computer Soc., Long Beach,
  CA, 1978, 48--54.

\bibitem[\protect\citename{McKay \& Radziszowski, }1997]{McKay97}
McKay, Brendan \& Radziszowski, Stanislaw. 1997.
\newblock Subgraph counting identities and {R}amsey numbers, to the fond memory
  of {Paul Erd\H{o}s}.
\newblock {\em Journal of Combinatorial Theory, Series B}, {\bf 69}, 193--209.
\newblock http://cs.anu.edu.au/$\sim$bdm/papers/r55.ps.gz.

\bibitem[\protect\citename{Mead {\em et~al.}, }1974]{Mead74}
Mead, Ernest; Rosa, Alexander; \& Huang, Charlotte. 1974.
\newblock The game of {SIM}: A winning strategy for the second player.
\newblock {\em Mathematics Magazine}, {\bf 47}(5), 243--247.

\bibitem[\protect\citename{Moore, }1987]{Moore87}
Moore, Thomas~E. 1987.
\newblock {SIM} on a microcomputer.
\newblock {\em J. Recreational Mathematics}, {\bf 19}(1), 25--29.

\bibitem[\protect\citename{Nairn \& Sperry, }1973]{Nairn73}
Nairn, John~H. \& Sperry, A.~B. 1973.
\newblock {SIM} on a desktop calculator.
\newblock {\em J. Recreational Mathematics}, {\bf 6}(4), 243--251.

\bibitem[\protect\citename{Ne\v{s}et\v{r}il, }1995]{Nesetril95}
Ne\v{s}et\v{r}il, Jaroslav. 1995.
\newblock Ramsey theory.
\newblock {\em Chap. 25, pages  1331--1403 of:} Graham, R.L.; Gr\"otschel, M.;
  \& Lov\'asz, L. (eds), {\em Handbook of Combinatorics}.
\newblock Elsevier.

\bibitem[\protect\citename{{O'Brian}, }1978-79]{OBrian78}
{O'Brian}, G.~L. 1978-79.
\newblock The graph of positions in the game of {SIM}.
\newblock {\em J. Recreational Mathematics}, {\bf 11}(1), 3--9.

\bibitem[\protect\citename{Papadimitriou, }1985]{Papadimitriou85}
Papadimitriou, Christos~H. 1985.
\newblock Games against {N}ature.
\newblock {\em Journal of Computer and System Sciences}, {\bf 31}, 288--301.

\bibitem[\protect\citename{Papadimitriou, }1994]{Papadimitriou94}
Papadimitriou, Christos~H. 1994.
\newblock {\em Computational complexity}.
\newblock Addison-Wesley.

\bibitem[\protect\citename{Peke\v{c}, }1996]{Pekec96}
Peke\v{c}, Aleksandar. 1996.
\newblock A winning strategy for the {R}amsey graph game.
\newblock {\em Combinatorics, Probability \& Computing}, {\bf 5}(3), 267--276.
\newblock http://dimacs.rutgers.edu/techps/1995/95-10.ps.gz.

\bibitem[\protect\citename{P\'olya, }1937]{Polya37}
P\'olya, George. 1937.
\newblock {Kombinatorische Anzahlbestimmungen f\"ur Gruppen, Graphen und
  chemische Verbindungen}.
\newblock {\em Acta Mathematica}, {\bf 68}, 145--254.
\newblock German. English translation by Dorethee Aeppli in: P\'olya, G. \&
  Read, R.~C. 1987. {\em Combinatorial Enumeration of Groups, Graphs and
  Chemical Compounds}. Springer Verlag, New York.

\bibitem[\protect\citename{Radziszowski, }1994; Revision \#6: July 5,
  1999]{Radziszowski94}
Radziszowski, Stanislaw. 1994; Revision \#6: July 5, 1999.
\newblock Dynamic surveys in combinatorics: Small ramsey numbers.
\newblock {\em Electronic Journal of Combinatorics}.
\newblock http://www.combinatorics.org/Surveys/ds1.ps.

\bibitem[\protect\citename{Ramsey, }1930]{Ramsey30}
Ramsey, Frank~P. 1930.
\newblock On a problem of formal logic.
\newblock {\em Proc.\ London Math.\ Soc.}, {\bf 30}(2), 264--286.
\newblock Reprinted in I. Gessel and G.-C. Rota, editors. Classic Papers in
  Combinatorics. Pages 2--24. Birkh\"auser Boston 1987.

\bibitem[\protect\citename{Richer, }1999]{Richer99}
Richer, Duncan~C. 1999.
\newblock {\em Maker-Breaker Games}.
\newblock Poster-presentation at Erd\H{o}s'99 July 4--11, 1999, in Budapest,
  Hungary.

\bibitem[\protect\citename{R\"odl, }1973]{Roedl73}
R\"odl, Vojtech. 1973.
\newblock {\em A generalization of {R}amsey theorem and dimension of graphs}.
\newblock MSc thesis, Charles University, Prague.

\bibitem[\protect\citename{Rounds \& Yau, }1974]{Rounds74}
Rounds, E.~M. \& Yau, S.~S. 1974.
\newblock A winning strategy for {SIM}.
\newblock {\em J. Recreational Mathematics}, {\bf 7}(3), 193--202.

\bibitem[\protect\citename{Schaefer, }1999]{Schaefer99}
Schaefer, Marcus. 1999 (May).
\newblock Graph {R}amsey theory and the polynomial hierarchy.
\newblock {\em In:} {\em Proc. 31st ACM Symposium on Theory of Computing
  (STOC)}.
\newblock SIGACT (the ACM Special Interest Group on Algorithms and Computation
  Theory), Atlanta, Georgia.
\newblock http://www.cs.uchicago.edu/publications/tech-reports/TR-98-04.ps.

\bibitem[\protect\citename{Schaefer, }1978]{Schaefer78}
Schaefer, Thomas~J. 1978.
\newblock On the complexity of some two-person perfect-information games.
\newblock {\em Journal of Computer and System Sciences}, {\bf 16}(2), 185--225.

\bibitem[\protect\citename{Schaeffer \& Lake, }1996]{Schaeffer96a}
Schaeffer, Jonathan \& Lake, Robert. 1996.
\newblock Solving the game of {C}heckers.
\newblock {\em Pages  119--133 of:} Nowakowski, Richard~J. (ed), {\em Games of
  No Chance}.
\newblock Mathematical Sciences Research Institute Publications, vol. 29.
\newblock Cambridge University Press.
\newblock
  http://www.msri.org/publications/books/Book29/files/schaeffer.$\{$ps.gz$|$pdf$\}$.

\bibitem[\protect\citename{Schwartz, }1979]{Schwartz79}
Schwartz, Benjamin~L. (ed). 1979.
\newblock {\em Mathematical solitaires and games}.
\newblock Farmingdale, NY: Baywood Publishing Company.
\newblock Pages  37--81.

\bibitem[\protect\citename{Schwartz, }1981-82]{Schwartz81}
Schwartz, Benjamin~L. 1981-82.
\newblock {SIM} with non-perfect players.
\newblock {\em J. Recreational Mathematics}, {\bf 14}(4), 261--265.

\bibitem[\protect\citename{Shader, }1978]{Shader78}
Shader, Leslie~E. 1978.
\newblock Another strategy for {SIM}.
\newblock {\em Mathematics Magazine}, {\bf 51}(1), 60--64.

\bibitem[\protect\citename{Shader \& Cook, }1980]{Shader80}
Shader, Leslie~E. \& Cook, Michael~L. 1980.
\newblock A winning strategy for the second player in the game {T}ri-tip.
\newblock {\em In:} {\em Proc.\ Tenth S.E.\ Conference on Computing,
  Combinatorics and Graph Theory}.

\bibitem[\protect\citename{Simmons, }1969]{Simmons69}
Simmons, Gustavus~J. 1969.
\newblock The game of {SIM}.
\newblock {\em J. Recreational Mathematics}, {\bf 2}(2), 66.

\bibitem[\protect\citename{Slany, }1988]{Slany88}
Slany, Wolfgang. 1988 (Jan.).
\newblock {\em {HEXI}: A happy perfect hexagone player}.
\newblock Tech. rept.~31. Institut f{\"u}r Statistik und Informatik der
  Universit\"at Wien.
\newblock MS-Dos executable:
  http://www.dbai.tuwien.ac.at/ftp/user/slany/hexi87.tgz.

\bibitem[\protect\citename{Thomsen \& {Leth Thomsen}, }1998]{Thomsen98}
Thomsen, Bent \& {Leth Thomsen}, Lone. 1998.
\newblock Towards global computations guided by concurrency theory.
\newblock {\em Bulletin of the European Association for Theoretical Computer
  Science}, Oct., 92--98.

\bibitem[\protect\citename{Tucker, }1995]{Tucker95}
Tucker, Alan. 1995.
\newblock {\em Applied combinatorics}. 3rd edn.
\newblock Wiley.

\bibitem[\protect\citename{Zhu, }1997]{Zhu97}
Zhu, Bidan. 1997.
\newblock {\em Ramsey theory illustrated through a {J}ava based game that plays
  heuristically and can learn in a client-server style or optimally if
  possible}.
\newblock Diplomarbeit ($\simeq$ MSc thesis), Technische Universit\"at Wien.
\newblock Java applet: http://www.dbai.tuwien.ac.at/proj/ramsey/.

\end{thebibliography}

\rem{[] noindents after theo etc, aber nicht zuviel!}

\rem{[] finally: check that no formulas span several pages} 

\rem{[] finally: figures at best places?}

\end{document}